\documentclass[english, norunningheads]{lni}

\usepackage[utf8]{inputenc}
\usepackage{listings}
\usepackage{listings-golang}
\usepackage{color}
\usepackage{tabularx}
\usepackage{tikz}
\usepackage{subcaption}

\lstdefinestyle{pseudo}{
  frame=tb,
  language={c++},
  deletekeywords={with},
  aboveskip=2mm,
  belowskip=2mm,
  captionpos=b,
  showstringspaces=false,
  columns=flexible,
  basicstyle={\footnotesize\ttfamily},
  numbers=left,
  numberstyle=\tiny \color{black},
  keywordstyle=\color{blue},
  commentstyle=\color{magenta},
  frame=none,
  breaklines=true,
  breakatwhitespace=true,
  tabsize=3,
}

\newcolumntype{L}[1]{>{\raggedright\let\newline\\\arraybackslash\hspace{0pt}}m{#1}}
\newcolumntype{C}[1]{>{\centering\let\newline\\\arraybackslash\hspace{0pt}}m{#1}}
\newcolumntype{R}[1]{>{\raggedleft\let\newline\\\arraybackslash\hspace{0pt}}m{#1}}

\newcommand*\circled[1]{\tikz[baseline=(char.base)]{
            \node[shape=circle,draw,inner sep=0.6pt] (char) {#1};}}

\begin{document}
\title[The Easiest Way of Turning your Relational Database into a Blockchain and the Cost of Doing So]{The Easiest Way of Turning your Relational Database into a Blockchain --- and the Cost of Doing So}
\author[Felix Schuhknecht \and Simon Jörz]
{Felix Schuhknecht\footnote{Johannes Gutenberg University Mainz, Institute of Computer Science, Staudingerweg 9, 55128 Mainz, 
Germany \email{schuhknecht@uni-mainz.de}} \and
Simon Jörz\footnote{Johannes Gutenberg University Mainz, Institute of Computer Science, Staudingerweg 9, 55128 Mainz, Germany
\email{sjoerz@students.uni-mainz.de}\newline}}
\startpage{1} 
\editor{GI} 
\booktitle{BTW 2023} 
\maketitle

\begin{abstract}
Blockchain systems essentially consist of two levels: The network level has the responsibility of distributing an ordered stream of transactions to all nodes of the network in exactly the same way, even in the presence of a certain amount of malicious parties (byzantine fault tolerance). On the node level, each node then receives this ordered stream of transactions and executes it within some sort of transaction processing system, typically to alter some kind of state.
This clear separation into two levels as well as drastically different application requirements have led to the materialization of the network level in form of so-called blockchain frameworks. While providing all the ``blockchain features'',  these frameworks leave the node level backend flexible or even left to be implemented depending on the specific needs of the application. 

In the following paper, we present how to integrate a highly versatile transaction processing system, namely a relational DBMS, into such a blockchain framework. As framework, we use the popular Tendermint Core, now part of the Ignite/Cosmos eco-system, which can run both public and permissioned networks and combine it with relational DBMSs as the backend. This results in a ``relational blockchain'', which is able to run deterministic SQL on a fully replicated relational database. 
Apart from presenting the integration and its pitfalls, we will carefully evaluate the performance implications of such combinations, in particular, the throughput and latency overhead caused by the blockchain layer on top of the DBMS. As a result, we give recommendations on how to run such a systems combination efficiently in practice.  
 
\end{abstract}


\section{Introduction}

In recent years, blockchain systems gained interest in various contexts, as they provide distributed transaction processing in potentially untrusted environments. Whereas the original applications mainly targeted public environments such as crypto currencies~\cite{bitcoin, ethereum}, blockchain systems have also gained interest in permissioned setups, where independent and potentially distrusting organizations, such as for instance companies trading with each other, want to perform some sort of mutual transaction processing~\cite{app1, app2, app3}. 

While the needs and environments for blockchain systems exist, a major downforce for the application of this technology has always been its hard entry level. Existing blockchain systems are often tailored towards a specific use-case or application domain and therefore are hard to apply for new application types. To deal with this challenge, one of the three following strategies is typically applied: (1)~To reinvent the wheel and to engineer a new blockchain system from scratch, fitting to the specific needs. (2)~To carefully adapt an existing blockchain system to the new requirements. (3)~To not install a blockchain solution at all. 
Of course, often, consequence~(3) is picked as (1)~and~(2) are cumbersome and therefore costly.

A step towards solving this problem is the observation that all blockchain systems essentially consist only of  two major components. 
The first component manages the network level. It receives input transactions, orders them globally, and distributes the transaction sequence to each node of the network in exactly the same way. The challenge here lies in performing this in an untrusted environment, where a certain amount of participants might behave maliciously or at least arbitrarily. To guarantee safety and liveness in such an environment, network levels implement sophisticated consensus mechanisms, secure message passing, and tamper-proof transaction logging. Despite various different implementations, the network level is rather independent from the actual application, as the semantics of the transactions are not relevant for this part.
The second component manages the node level and centers around the processing of transactions within each node. Naturally, the requirements here are highly application dependent. For instance, to implement a currency, a simple key-value-store backend managing account balances with a \texttt{put()}/\texttt{get()}/\texttt{delete()} interface is sufficient. However, to power complex sales operations, a relational database system offering full-fledged SQL support is certainly the better choice. 

As a consequence of these observations, blockchain frameworks have emerged that try to strictly separate their components by design. For example, the popular permissioned blockchain framework Hyperledger Fabric~\cite{fabric} provides exchangeable default implementations for its individual components. While this is a step in the right direction, Fabric unfortunately still forces the user to stick to the key-value model. This limitation is overcome by the framework Tendermint Core~\cite{tendermint}, which leaves the node level backend fully unimplemented. It is up the application to provide a backend which receives and processes the transactions that are distributed by the framework to each node.

As this is a promising design for tackling the initial problem, in this work, we investigate how challenging and how practical it is to connect a transaction processing backend to the blockchain framework Tendermint Core. To cover a wide range of applications, we integrate a highly-versatile transaction processing system, namely a relational DBMS. We show how to create a ``relational blockchain'' with minimal effort and are especially interested in the overhead that is caused by this combination. We will investigate the latency and throughput of the relational blockchain under the drastically different synchronous, pseudo-synchronous, and asynchronous communication, each appropriate for different types of applications. Further, we will look at the scaling behavior of the system and discuss important configuration parameters. In summary, we will provide recommendations on how to use such a relational blockchain efficiently in practice. 

\subsection{Contributions}

\begin{enumerate}
\item We present how to integrate a stand-alone single-node relational DBMS into the blockchain framework Tendermint. Our current implementation supports PostgreSQL and MySQL and can easily be extended for further systems. As a result of this combination, we produce a \textit{relational blockchain} that can execute (deterministic) SQL transactions equally across a set of potentially untrusted nodes to modify a fully replicated database.
\item We evaluate \textit{latency} and \textit{throughput/end-to-end runtime} of the relational blockchain under Smallbank~\cite{smallbank} and TPC-C~\cite{tpcc} transactions. We compare its performance with a standalone execution of the workloads in PostgreSQL to identify the overhead that is caused by the blockchain framework on top of the relational backend. 
\item We evaluate the impact of three different \textit{communication methods}, namely synchronous, pseudo-synchronous, and asynchronous communication. We show that the choice of the communication method has a drastic impact on the performance of the system.   
\item We evaluate the impact of the \textit{relational backends}, namely PostgreSQL and MySQL, under synchronous and asynchronous communication.
\item We evaluate the \textit{scaling capabilities} of the relational blockchain. Here, we first scale the number of virtual nodes within a physical node, which factors out network latency and resembles the Blockchain-as-a-Service (BaaS) setup. Then, we scale number of physical nodes within and across data-centers, resembling the classical distributed setup, facing network/internet latency.
\item We provide \textit{practical recommendations} in which situations a relational blockchain yields a good performance -- and in which situations it does not. To allow and easy application of our findings, we will release all code, results, scripts and auxiliary material of this paper in the repository: \href{https://gitlab.rlp.net/fschuhkn/relational-blockchain}{https://gitlab.rlp.net/fschuhkn/relational-blockchain}
\end{enumerate}

The paper is structured as follows: In Section~\ref{sec:related_work}, we start with a discussion of the related work in the field. In Section~\ref{sec:relational_blockchain}, we present how to integrate the support for relational DBMSs into Tendermint Core. Therein, we also describe the differences between the different types of communication. In Section~\ref{sec:experimental_setup}, we discuss the precise setup and empirically determined configuration used for the evaluation. In Section~\ref{sec:experimental_evaluation}, we perform an extensive experimental evaluation of the relational blockchain with a focus on the generated overhead and scaling capabilities. Finally, in Section~\ref{sec:conclusion}, we conclude with recommendations on whether and how to use such a systems combination in practice.

\section{Related Work}
\label{sec:related_work}

Before presenting our relational blockchain, let us discuss other work that sits at the intersection of blockchains and database systems. 

There exists other interesting work that analyzes and/or builds upon the Tendermint framework. In~\cite{tendermint_rel1, tendermint_rel2}, the authors perform an interesting performance analysis of the internal behavior of the framework. In~\cite{tendermint_rel3}, the authors analyze correctness and fairness of the system. The findings in these works justify our use and setup of Tendermint: The framework powers hundreds of applications of the Cosmos network, where most networks are tightly coupled with only few nodes. Latency and throughput decreases gracefully with the number of nodes participating in the consensus. 
Tendermint has also been used before to connect DBMSs as the backend. A prominent example is BigchainDB~\cite{bigchaindb}, which uses the document store MongoDB~\cite{mongodb} as backend.

Apart from Tendermint, there exist other blockchain frameworks. The most prominent representative is clearly Hyperledger Fabric~\cite{fabric}, designed to power permissioned blockchain networks. The modular design is composed of interchangeable components that allow a tuning of the network to the specific needs of the application up to a certain degree. Unfortunately, the system is hardcoded against a key-value model, such that the integration of a relational backend is not possible without deep changes of the system. Another blockchain framework is ChainifyDB~\cite{chainifydb}, that allows the creation of heterogeneous blockchain networks. Here, heterogeneous means that different relational systems can be used across a single network. The applied processing model still ensures correctness of transaction processing.      

Apart from frameworks, many research papers discuss the interconnection and relation of classical DBMSs and blockchain systems and how to combine both worlds. In BlockchainDB~\cite{blockchaindb1, blockchaindb2}, a database layer is placed on top of a blockchain layer to combine the proper query interface of a database systems with the replication guarantees of a blockchain. 
In~\cite{blockchain_meets_db}, the authors take the other route and extend a relational system, namely PostgreSQL, with a blockchain layer in order to create a blockchain network between multiple PostgreSQL instances. Unfortunately, this project requires a deep modification of PostgreSQL.
Another interesting project is Veritas~\cite{veritas}. Therein, the authors propose to extend existing DBMSs with blockchain features in a cloud environment. 

Apart from architectural works, many projects try to improve the performance of blockchain systems in order to converge towards the performance of traditional (distributed) DBMSs. In Fabric++~\cite{fabric++}, several optimization techniques from the database domain are transferred to Fabric in order to speed up processing.
Other works try to improve blockchain performance via sharding~\cite{sharding} and various low-level optimizations in the transaction processing flow~\cite{fastfabric}.

\section{Setting up a Relational Blockchain}
\label{sec:relational_blockchain}

In the following section, we will discuss how to integrate a relational DBMS into the Tendermint framework, which we believe is a good template for how blockchain frameworks are reasonably engineered. On the backend side, we will focus on relational DBMSs in this work. However, the general process is applicable to non-relational transaction processing backends in a similar fashion. 

\subsection{The Blockchain Framework: Tendermint Core}

The design goal of the blockchain framework Tendermint Core~\cite{tendermint} is to provide essentially all those components that are shared in typical blockchain environments~\cite{untangling_blockchain, talking_blockchains}, but nothing more than that. Precisely, the entire transaction processing backend is left unimplemented and must be provided by the application side. There are two requirements for the backend: (1)~The same backend must be used within all nodes of the network. (2)~This backend must be deterministic, i.e., it executes a block of transactions in the same way on all nodes. 
Overall, this design has pleasant advantages: 
On one hand, such a framework approach removes the need to reinvent the wheel by reimplementing essential components of blockchain systems. On the other hand, it grants the framework flexibility to support a large variety of applications.  

The most essential components that are already provided by Tendermint Core are:
\begin{enumerate}
\item A \textit{transaction pool} which has the responsibility to receive and hold transactions that are pending for ordering and execution. All submitted input transactions first go into this pool, where they can be rejected already, if they do not match user-specified criteria, by implementing the function~\texttt{CheckTx()}. The pool itself is lazily replicated across the nodes, i.e., nodes share pending transactions with other nodes via gossip broadcasting.
\item A \textit{consensus mechanism} called Polka, which is a variation of the well-known PBFT~\cite{pbft} consensus. It can tolerate up to $f$ maliciously behaving parties in a set of $3f+1$ parties in total. While the mechanism is tailored towards a permissioned setup, where all participants are known at all times, it can be extended to work in a public environment as well by using a Proof-of-Stake-like approach. As this requires the integration of a currency, in this work, we run the default version of the consensus mechanism in a permissioned environment.  
\item The \textit{ledger}, which stores the observed sequence of committed transactions at the granularity of blocks within each node in a tamper-resistant way.   
\item A \textit{message passing system} that ensures a secure communication between individual parties of the network.

\end{enumerate} 

On the network level, the workflow of the system essentially looks as follows: First, a client submits a new transaction to the network. The network then stores this transaction in the transaction pool with other pending transactions. A node then picks a set of transactions from the pool and groups them into a block in an ordered way. The block then goes through multiple consensus rounds until it is either globally decided that this block will be accepted or it is globally rejected. If it is rejected, another block will be proposed (potentially by another node) and consensus restarts. However, if the block is accepted, it is distributed to all nodes of the network. Each node that receives a block then appends it to its copy of the ledger and passes the block to the transaction processing backend.

\subsection{Communicating with the Transaction Processing Backend}
\label{ssec:backend_communcation}

The block passing between the framework and the transaction processing backend happens via a so-called \textit{Application Blockchain Interface (ABCI)}. The interface essentially consists only of the four functions \texttt{BeginBlock()}, \texttt{DeliverTx()}, \texttt{EndBlock()}, and \texttt{Commit()}, which must be implemented by the backend and which are called by the framework. For every agreed-upon block that is distributed, the core first calls \texttt{BeginBlock()} on each node to signal the arrival of a new block to the backend. Then, for each transaction within the block, the core calls \texttt{DeliverTx()} sequentially. This function is responsible for the actual processing of the transaction. It also returns whether the execution of a transaction was successful or not. After all transactions have been delivered, the core calls \texttt{EndBlock()} to signal that the block is done. Finally, the core calls \texttt{Commit()}. This tells the backend that all changes made by the transactions of the block must become real and visible for upcoming processing, if all transactions in the block succeeded. Otherwise, \texttt{Commit()} is responsible for rolling back all changes made by all transactions of the block. 

To implement this ABCI and to connect a backend to Tendermint core, there are two options which we call the \textit{server-variant} and the \textit{builtin-variant}. In the server-variant, the backend implementing the ABCI runs as an independent socket-server and the core calls the interface via TCP. In the builtin-variant, the ABCI is implemented by the backend as a component of Tendermint core and directly compiled into it. While the server-variant offers a higher flexibility, the builtin-variant allows the core to communicate with the backend via simple function calls. In Section~\ref{ssec:config}, we will evaluate both variants.

\subsection{Integrating a Relational DBMS as Backend}

Connecting a relational DBMS to the blockchain framework by implementing the ABCI is fairly natural, as both sides provide transaction semantics. However, to avoid confusion, we now have to clearly differentiate between different types of transactions and different types of commits in our system composition: We will call transactions, that are submitted to the blockchain network as \textit{bc-transactions}. As discussed, multiple bc-transactions can be grouped in a block, which is committed as a whole by the framework. We call this a \textit{bc-commit}. In contrast to that, we refer to transactions that are executed by the relational DBMS as \textit{db-transactions}. The DBMS commits at the granularity of individual db-transactions, which we call \textit{db-commit}.

Before being able to communicate with the relational DBMS from within the ABCI functions, we establish a connection to it in the bootstrapping part of Tendermint Core. To do so, we utilize the drivers \texttt{pgx}~\cite{pgx} and \texttt{go-sql-driver/mysql}~\cite{mysql-driver} for PostgreSQL and MySQL, respectively, to open a connection to the DBMS instance.  
Table~\ref{tab:abci} now shows the pseudo-code\footnote{The actual implementation of the ABCI in Go is actually only marginally more complex.} implementation of \texttt{BeginBlock()}, \texttt{DeliverTx()}, and \texttt{Commit()}, where we show only the communication with the DBMS and removed any boilerplate code or error handling. As \texttt{EndBlock()} does not involve any DBMS communication, we do not show it here.

\begin{table}
\begin{tabular}{ L{.27\linewidth} L{.36\linewidth}  L{.33\linewidth}}
\hspace{-0.3cm}
\begin{minipage}[t]{\linewidth}
\begin{small}
\begin{lstlisting}[style=pseudo, label={listing:abci_beginblock}, xleftmargin=0.0ex, escapechar=|]
BeginBlock() {
  // start db-transaction
  db-transaction dbTx 
    = db.Begin()
  return dbTx
}
\end{lstlisting}
\end{small}
\end{minipage}
&
\begin{minipage}[t]{\linewidth}
\begin{small}
\begin{lstlisting}[style=pseudo, label={listing:abci_delivertx}, xleftmargin=0.0ex, escapechar=|]
DeliverTx(db-transaction dbTx,
          bc-transaction bsTx) {
  // extract SQL statement 
  // from bc-transaction
  sql stmt = DecodeTx(bsTx)
  // execute SQL statement 
  // as part of db-transaction
  status s = dbTx.Execute(stmt)
  return s
}
\end{lstlisting}
\end{small}
\end{minipage}
&
\begin{minipage}[t]{\linewidth}
\begin{small}
\begin{lstlisting}[style=pseudo, label={listing:abci_commit}, xleftmargin=0.0ex, escapechar=|]
Commit(db-transaction dbTx,
       status[] s) {
  if(s.Contains(bsTxFailed))
    dbTx.Rollback()
  else
    // perform db-commit 
    // (= perform bc-commit)
    dbTx.Commit()
}
\end{lstlisting}
\end{small}
\end{minipage}
\end{tabular}
\caption{Pseudo-code for \texttt{BeginBlock()}, \texttt{DeliverTx()}, and \texttt{Commit()} that focuses on the communication with the DBMS.}
\label{tab:abci}
\end{table}

In our implementation, \texttt{BeginBlock()} has the sole purpose to begin a new db-transaction. The context \texttt{db} is provided by Tendermint and implements a generic interface from the Go package \texttt{sql}~\cite{package_sql} that allows the communication with relational DBMSs. Relying on a generic interface enables an easy switching between PostgreSQL and MySQL (and other relational systems). Underneath this generic interface, we again use \texttt{pgx} respectively \texttt{go-sql-driver/mysql} as a compatibility layer. It essentially translates the generic calls to their DBMS-specific counterparts. 

In each call to \texttt{DeliverTx()}, we receive the db-transaction in progress as well as a bc-transaction of the current block. We first decode the received bc-transaction and extract the SQL statement that is stored therein as a string. Then, we pass the SQL statement to the db-transaction context for execution. This execution returns a status (success or failure), which also contains the result of the db-transaction. We return this status to Tendermint Core.

After all bc-transactions have been delivered, \texttt{Commit()} is called, which receives the open db-transaction and the execution statuses of all bc-transactions of the block. Based on the statuses, we check whether there is a bc-transaction that failed the execution. This could for instance be the case if the SQL statement contained in a bc-transaction is malformed\footnote{This could be checked as well in \texttt{CheckTx()} that is executed when a bc-transaction is inserted into the transaction pool, such that malformed bc-transactions do not become part of a block in the first place. However, since we do not inject malformed bc-transactions in our runs, this is currently not part of our implementation.}. 
If a failed bc-transaction exists, we command the DBMS to rollback the db-transaction, including all changes made by bc-transactions of the block. Otherwise, we can safely db-commit the db-transaction, such that all changes of this block become visible for the processing of the next block.  

Note that we use the previously described communication protocol only for \textit{modifying} transactions. To answer read-only transactions, we implement the ABCI function \texttt{Query()} which allows us to fire read-only queries against the backend of a single node\footnote{This can be extended to query multiple nodes to handle the risk of querying a byzantine node.}, effectively bypassing the costly transaction processing flow of the blockchain framework. 

\subsection{Synchronous vs Asynchronous Transaction Processing}
\label{ssec:communication}

To process modifying transactions in the blockchain network, the client has essentially two different modes available: (1)~A \textit{synchronous mode}, where a client-request blocks until it receives an answer from the system. (2)~An \textit{asynchronous mode}, where the request returns before receiving an answer. As we will evaluate both modes in the following, let us discuss their precise realization and behavior in the following.

We start with synchronous processing. First of all, to communicate with Tendermint Core, the client uses a Broadcast API in order to submit bc-transactions. From this API, we utilize the function \texttt{BroadcastTxCommit()}. This function basically resembles a synchronous submit that receives a bc-transaction and blocks until either it has been worked into a bc-committed block or it is rejected from the transaction pool (due to being malformed).
Consequently, our test suite looks fairly simple for the synchronous case: In each iteration of the loop, a client fires a bc-transaction using \texttt{BroadcastTxCommit()} and waits for the result before proceeding with the next iteration.

\begin{figure}[h]
\centering
\includegraphics[page=1, width=.95\linewidth, trim={0 13cm 4.9cm 0}, clip]{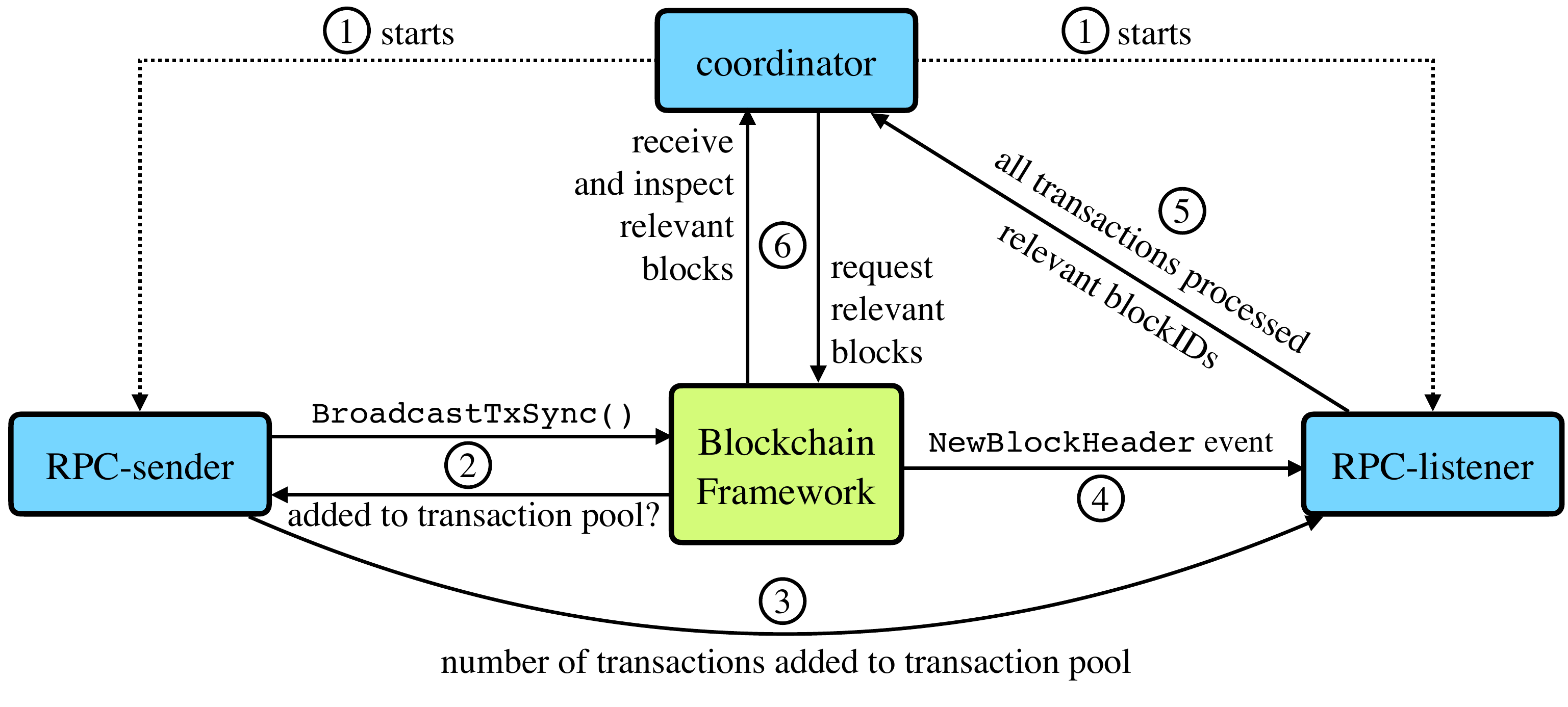}
\caption{Workflow of asynchronous transactions processing.}
\label{fig:async_test_suite}
\end{figure}

The asynchronous transaction processing is more complex. Here, we use the weaker API function \texttt{BroadcastTxSync()} to communicate with Tendermint Core, which already returns after the bc-transaction has been successfully\footnote{There is an even weaker function called \texttt{BroadcastTxAsync()}, which returns immediately after submission. However, as this function does not provide any information on whether the bc-transaction was submitted successfully, we do not consider it further.} added to the transaction pool. Thus, the client does not get a synchronous response on whether the transaction was committed successfully in a block or not. 
As we still require a reliable feedback on the success of execution, we implement a test suite as depicted in Figure~\ref{fig:async_test_suite}.   

It consists of three components: (a)~The \textit{coordinator}, responsible for orchestrating the entire run. (b)~The \textit{RPC-sender}, which broadcasts the bc-transactions to the framework. (c)~The \textit{RPC-listener}, which listens for bc-committed blocks.  
In \circled{1}, the main loop first starts both RPC-sender and RPC-listener. Then, in \circled{2}, the RPC-sender uses the aforementioned \texttt{BroadcastTxSync()} to push bc-transactions into the network. While doing so, the RPC-sender monitors the number of bc-transactions that made it into the transaction pool -- this is the number of transactions expected to make it through the system. In \circled{3}, after submitting all bc-transactions, this number is passed to the RPC-listener. For every block that is bc-committed by the framework, in \circled{4}, the RPC-listener receives a \texttt{NewBlockHeader} event from the framework and calculates the number of already seen bc-transactions based on it. As soon as it has seen all previously entered bc-transactions, in \circled{5}, it informs the coordinator that all bc-transactions have now been processed and passes the blockIDs containing these transactions. In \circled{6}, the coordinator then requests all relevant blocks and checks whether the bc-transactions have been processed successfully.  Note that the steps \circled{2} and \circled{4} can happen interleaved, i.e., the sender can still push in new bc-transactions while the listener is already receiving headers of committed blocks.

\subsection{Deterministic Execution}

As the blockchain framework essentially resembles a fully replicated state machine, it requires the backend to behave deterministically. We ensure a deterministic execution in two steps: 
(1)~The repetitive calls to the ABCI function \texttt{DeliverTx()} by the framework happen sequentially. As we ensure that any communication with the relational DBMS within \texttt{DeliverTx()} happens synchronously, all bc-transactions will be executed within a db-transaction in exactly the same order within the relational DBMS of each node. 
(2)~We submit only bc-transactions that contain deterministic SQL statements.

\section{Experimental Setup}
\label{sec:experimental_setup}

Before starting with our actual experimental evaluation and analysis, let us discuss the setup in the following. 
We start with a description of the test systems (Section~\ref{ssec:systems}) and the benchmarks (Section~\ref{ssec:benchmarks}). Then, we describe the setup of the framework and the backend (Section~\ref{ssec:config}). This includes making reasonable decisions for certain configurations types and parameters, which we evaluate empirically in the following. 

\subsection{Systems}
\label{ssec:systems}

As blockchain systems are used in different setups, we will evaluate different network configurations.  

First, to completely factor out network latency overhead, we perform a set of experiments on a single powerful machine equipped with an Intel i9-12900K CPU (Alder Lake) running at up to 5.2GHz with 16 cores. 
This state-of-the-art processor is able to run a set of virtual nodes and simulates a very low latency blockchain network. The machine contains 128GB of DDR4-3200. All database files are located on a 2TB Samsung 980 Pro M2 PCIe~4.0 SSD. As operating system, a 64-bit Arch Linux is installed. Note that such a setup consisting of a single physical node running the blockchain network is not fully artificial nowadays: So called Blockchain-as-a-Service (BaaS) solutions~\cite{baas, baas2} host all virtual nodes of the network in a single data-center, often also on the same physical node.   

Second, to measure the impact of a distributed setup across the internet, we also perform an additional set of experiments on a network of up to eight AWS EC2 instances (t2.small), which are distributed across the four regions Frankfurt, Ireland, London, and Paris, with up to two instances per region. Each instance has one~vCPU, contains $2$GB of RAM, has $16$GB of gp2 volume attached (general purpose SSD), and runs Ubuntu 20.04. Additionally, in the Frankfurt data-center, we run a separate instance (t2.micro) that serves as the client and orchestrates our runs.

\subsection{Benchmarks: Smallbank \& TPC-C}
\label{ssec:benchmarks}

In the following evaluation, we use transactions and datasets from two established transactional benchmarks from the world of blockchains and databases, namely Smallbank~\cite{smallbank} and TPC-C~\cite{tpcc}. We use transactions from these two benchmarks as they offer very different characteristics: While Smallbank contains a set of five extremely simple and short-running transactions which essentially resemble only money transfers between accounts, the three used TPC-C transactions are far more complex and long-running. In the following, we give a brief overview of the used benchmarks.

For Smallbank, the database consists of a single table with four columns, where each tuple contains a user-ID and a name having both a balance for a checking account and a savings account, initialized with random integers. We use the five modifying transactions that are specified in the original benchmark description. The transactions~\texttt{TransactSavings} and \texttt{DepositChecking} each increase the respective account balance. \texttt{SendPayment} modifies two checking account balances. \texttt{WriteCheck} decreases a checking account balance. Finally, \texttt{Amalgamate} moves money from a savings account to the checking account of the same user. For each transaction, we randomly pick the account(s) as well as the amount to modify/move following a uniform distribution. 

For TPC-C, the database has nine tables in total and essentially represents a multi-warehouse wholesale operation. We implement the two modifying transactions \texttt{NewOrder} and  \texttt{Payment} and select the parameters of each fired transaction randomly within meaningful bounds as specified by the TPC-C benchmark description. Additionally, we implement the read-only transaction \texttt{OrderStatus} to test the query-interface of the framework.
We selected these three transactions as they are rather complex by accessing all nine tables and by modifying five of them, resulting in more long-running transactions than for Smallbank. The warehouses and districts are accessed by the transactions following a uniform distribution. 

Note that all Smallbank transactions are transmitted to the system on-the-fly. In contrast to that, the TPC-C transactions are registered in the relational DBMS as stored procedures due to their significantly higher code complexity and size. The transactions of TPC-C then simply contain a call of the corresponding stored procedure.

\subsection{Framework and Backend Configuration}
\label{ssec:config}

For Tendermint Core, we use the latest stable version 0.34 for all experiments. For PostgreSQL, we use version~14.5, for MySQL, we run version~8.0.30. Tendermint Core, PostgreSQL, and MySQL run in Docker containers and are installed from the corresponding Docker images. 

\begin{figure}[h!]
  \centering
  \begin{subfigure}[b]{.49\textwidth}
	\includegraphics[width=\linewidth, height=4.3cm, trim={0 0 0 1.1cm}, clip]{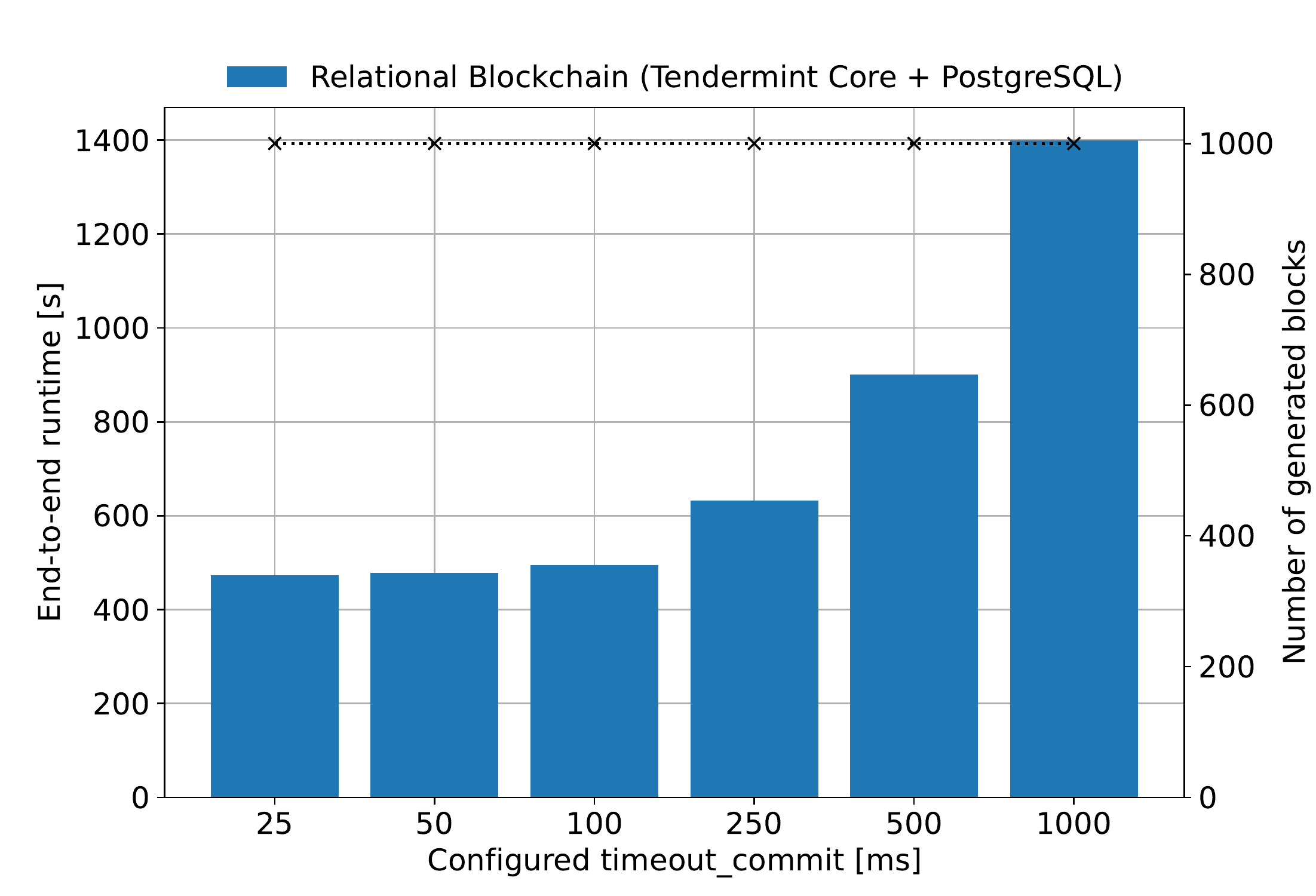}
    \caption{Synchronous communication.}
    \label{fig:timeout_commit_sbm_sync}
  \end{subfigure}
  \begin{subfigure}[b]{.49\textwidth}
	\includegraphics[width=\linewidth, trim={0 0 0 1.1cm}, clip]{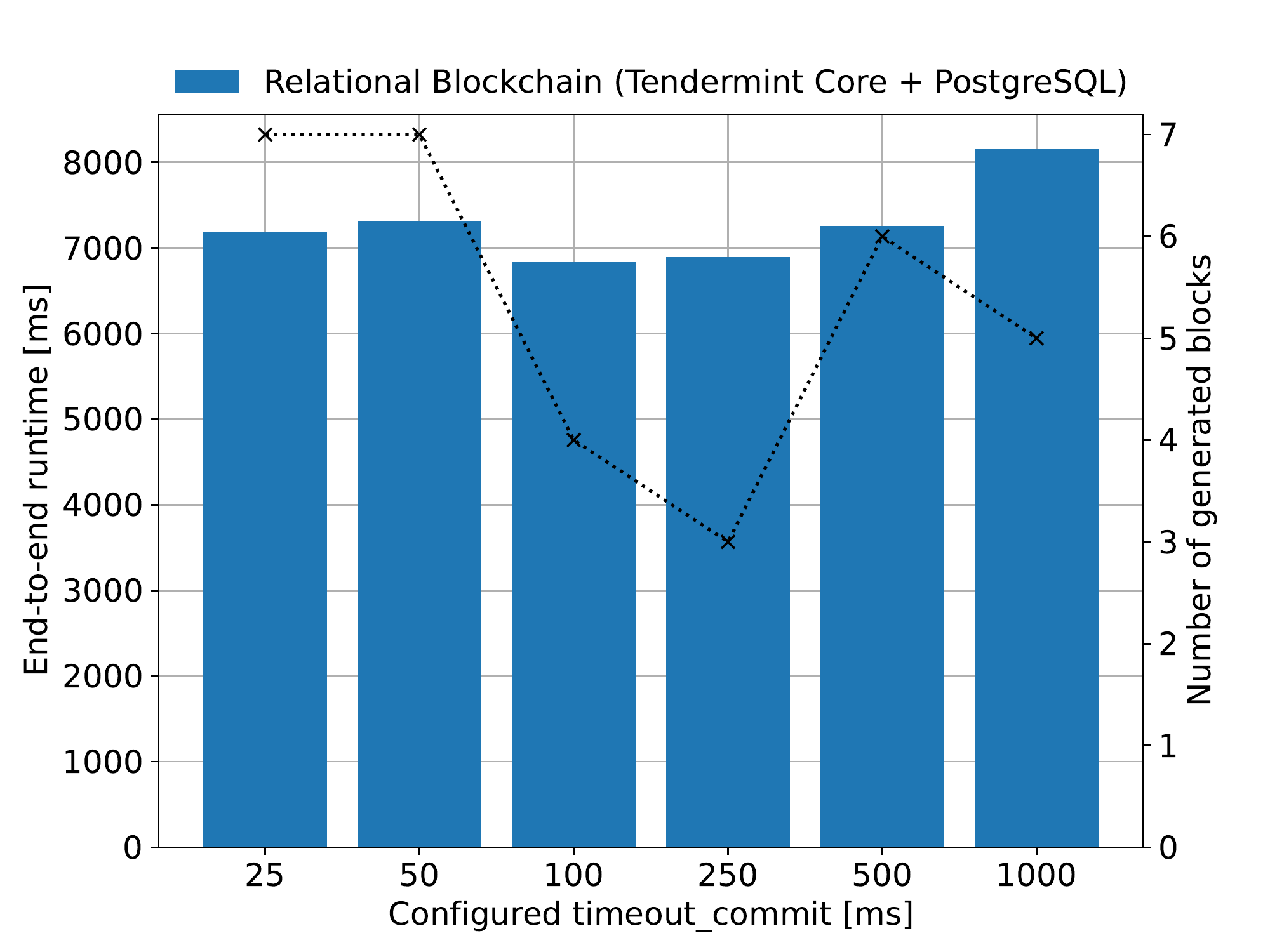}
    \caption{Asynchronous communication.}
    \label{fig:timeout_commit_sbm_async}
  \end{subfigure}
  \caption{Varying the \texttt{timeout\_commit} parameter from $25$ms to $1000$ms.}
  \label{fig:timeout_commit_smb}
\end{figure}

Tendermint Core as well as the relational DBMS are deeply configurable. To measure the ``out-of-the-box'' performance, we start with the default configuration and try to tune the systems as little as possible. We state and justify in the following all made changes:

On the side of Tendermint Core, we first increase the size of the transaction pool from $5{,}000$ to $100{,}000$ transactions, such that the whole transaction sequence of each benchmark always fits in. Next, there are multiple timeout parameters that have an impact on both the performance and the behavior of the network. We set \texttt{timeout\_broadcast\_tx\_commit} to a sufficiently large value ($10$s), such that synchronous communication never times out in our experiments. Also, we have to tune the important parameter \texttt{timeout\_commit}, which determines how long the consensus mechanism does wait for additional votes, if $2/3$~of the votes have been received already. To empirically identify a good value, in Figure~\ref{fig:timeout_commit_smb}, we perform an experiment where we vary \texttt{timeout\_commit} from $25$ms to $1000$ms for both $1{,}000$~synchronous and $10{,}000$~asynchronous transactions of Smallbank. As a value of $100$ms yields the best end-to-end runtime in both cases, we use a timeout of~$100$ms in all upcoming experiments. 
Further, we set the maximum allowed block size to $21$MB such that the size is never the limiting factor for forming a block. We disable the creation of empty blocks (in case the transaction pool runs dry) as well. 

For PostgreSQL and MySQL, we essentially keep the configuration of the used Docker images as is.

\begin{figure}[h!]
  \centering
  \begin{subfigure}[b]{.49\textwidth}
      \includegraphics[page=1, width=\linewidth, trim={0 0 2cm 0}, clip]{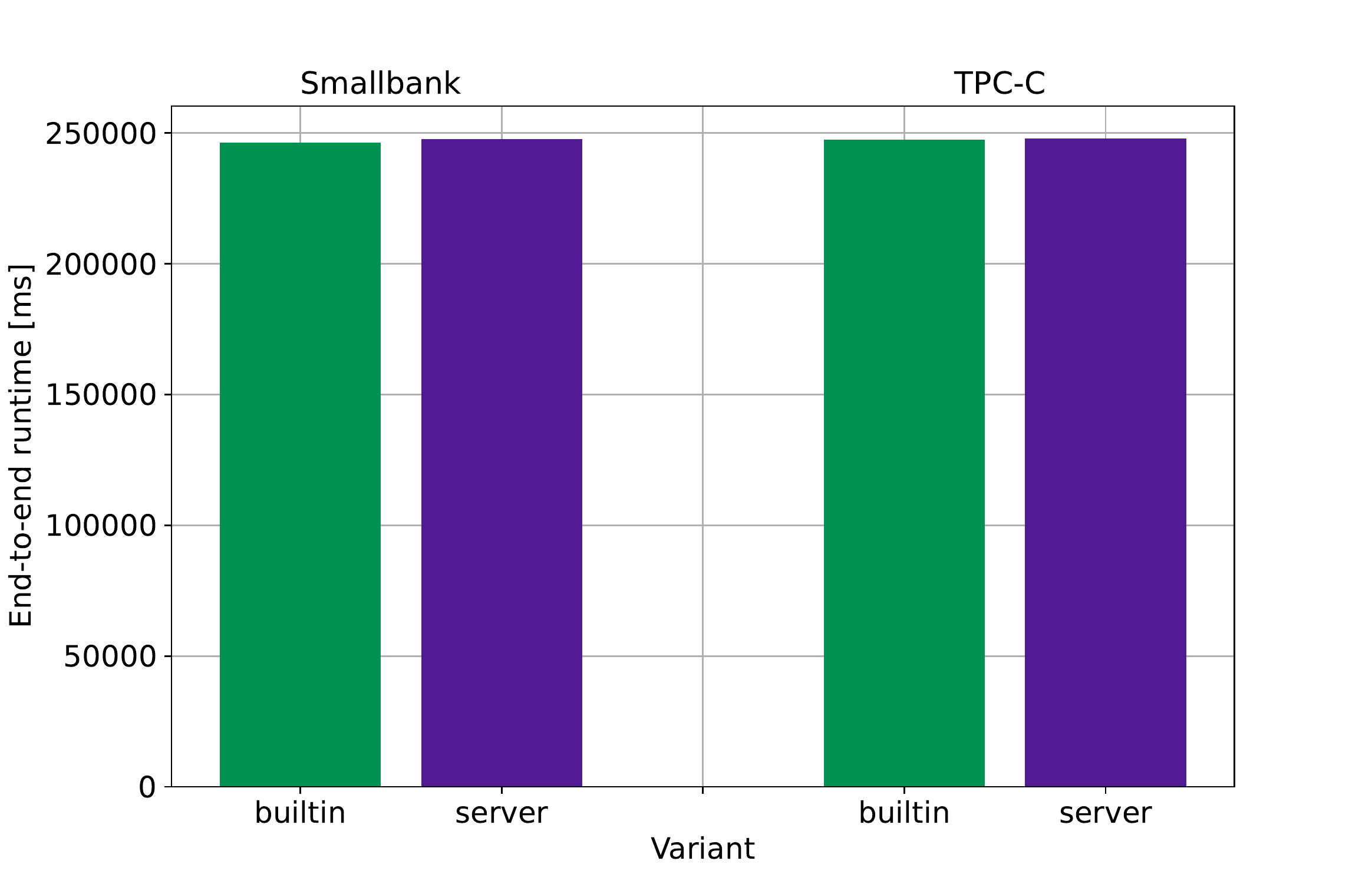}
    \caption{Synchronous communication.}
    \label{fig:server_vs_builtin_sync}
  \end{subfigure}
  \begin{subfigure}[b]{.49\textwidth}
  \includegraphics[page=1, width=.85\linewidth, trim={0 0 2cm 0}, clip]{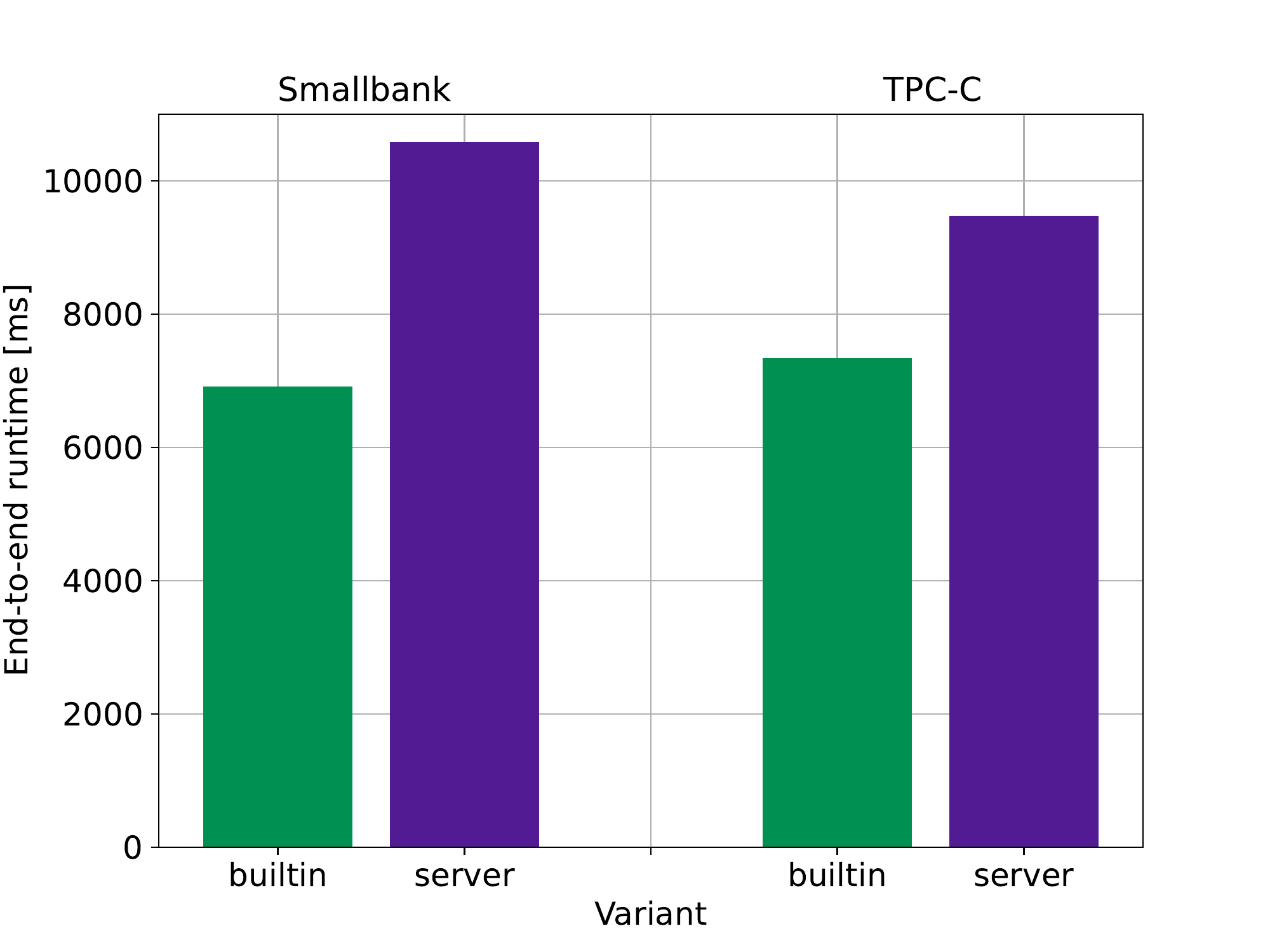}
    \caption{Asynchronous communication.}
    \label{fig:server_vs_builtin_async}
  \end{subfigure}
  \caption{Server-variant vs builtin-variant.}
  \label{fig:server_vs_builtin}
\end{figure}

As mentioned in Section~\ref{ssec:backend_communcation}, there are two ways of connecting the backend to the framework, where the server-variant is more flexible than the builtin-variant. To identify the performance impact, we implemented both variants and evaluate them against Smallbank and TPC-C transactions. Figure~\ref{fig:server_vs_builtin} shows the results for both synchronous and asynchronous communication.

We can see that for synchronous communication, there is hardly a difference visible. The type of connecting the backend is fully overshadowed by the high cost of synchronous communication (which we will evaluate in detail in Section~\ref{ssec:synchronous}). However, for the cheaper asynchronous communication, the builtin-variant is $1.53$x faster for Smallbank and  $1.23$x faster for TPC-C than the server-variant. This is due to the fact that in the server-variant the ABCI calls happen via TPC sockets, which are significantly more expensive than the direct function calls in the builtin-variant. Due to the higher performance, we use the builtin-variant in all following experiments.

\section{Experimental Evaluation \& Analysis}
\label{sec:experimental_evaluation}

In the following, we perform a set of experiments to determine the performance of the relational blockchain. We are particularly interested in its overhead (Section~\ref{ssec:overhead}) over the raw relational DBMS under synchronous communication (Section~\ref{ssec:synchronous}) and asynchronous communication (Section~\ref{ssec:asynchronous}). Then, we investigate the impact of the relational DBMS (Section~\ref{ssec:dbms}) and the scaling capabilities of the network (Section~\ref{ssec:scaling}).

\subsection{Overhead of the Blockchain Framework}
\label{ssec:overhead}

We start our experimental evaluation with the central question of how much overhead the blockchain frameworks actually adds on top of the relational DBMS. To see the overhead, we compare the performance of our relational blockchain setup (Tendermint Core + PostgreSQL) as described previously, with the performance of the raw DBMS (PostgreSQL only).

As setup for this experiment, we use a fairly typical permissioned configuration: We have a single client firing the bc-transactions into a network of four virtual nodes, where each node consists of a Tendermint Core instance as well as a PostgreSQL instance, each running in its docker container. Again, we use virtual nodes here to factor out any network latency. We test two workloads: (1)~The previously described TPC-C workload with $10$~warehouses. (2)~The Smallbank workload with $100{,}000$~accounts. Note that to distinguish the transactions of the workload from bc-transactions/db-transaction, we will call the former \textit{wl-transactions} in the following.  
As the type of communication plays a drastic role for both cost and usability, we perform in the following a synchronous as well as an asynchronous variant of the experiment.  

\subsubsection{Synchronous and Pseudo-synchronous Communication}
\label{ssec:synchronous}

We start with synchronous communication. It basically resembles the typical communication with a DBMS: A client submits a transaction to the system and the call blocks until eventually, it returns the result.  
As we have described in Section~\ref{ssec:communication}, the blockchain framework supports such a communication style via its Broadcast-API. 
Additional to this fully synchronous communication, where one wl-transaction is packed into one bc-transaction, we also test a pseudo-synchronous communication style. Therein, we pack multiple wl-transactions into a single bc-transaction and fire this bc-transaction synchronously. On one hand, this results in less bc-transactions that have to go through the system, potentially lowering the pressure on the network and the transaction processing overhead. On the other hand, this relaxes our notion of synchronicity (hence pseudo), as the client receives a synchronous response only for a batch of wl-transactions, not for each wl-transaction individually.   

To asses the overhead, we are interested in both the \textit{latency} and the \textit{end-to-end runtime}. In this context, latency is the time between submitting a bc-transaction and receiving a response to it. Note that we submit a new bc-transaction only after receiving a response to the previous one. The end-to-end runtime is the time between submitting the first bc-transaction and receiving the response to the last bc-transaction.

\begin{figure}[h!]
  \centering
  \begin{subfigure}[b]{.49\textwidth}
	\includegraphics[width=\linewidth, trim={0.3cm 0 2cm 0}, clip]{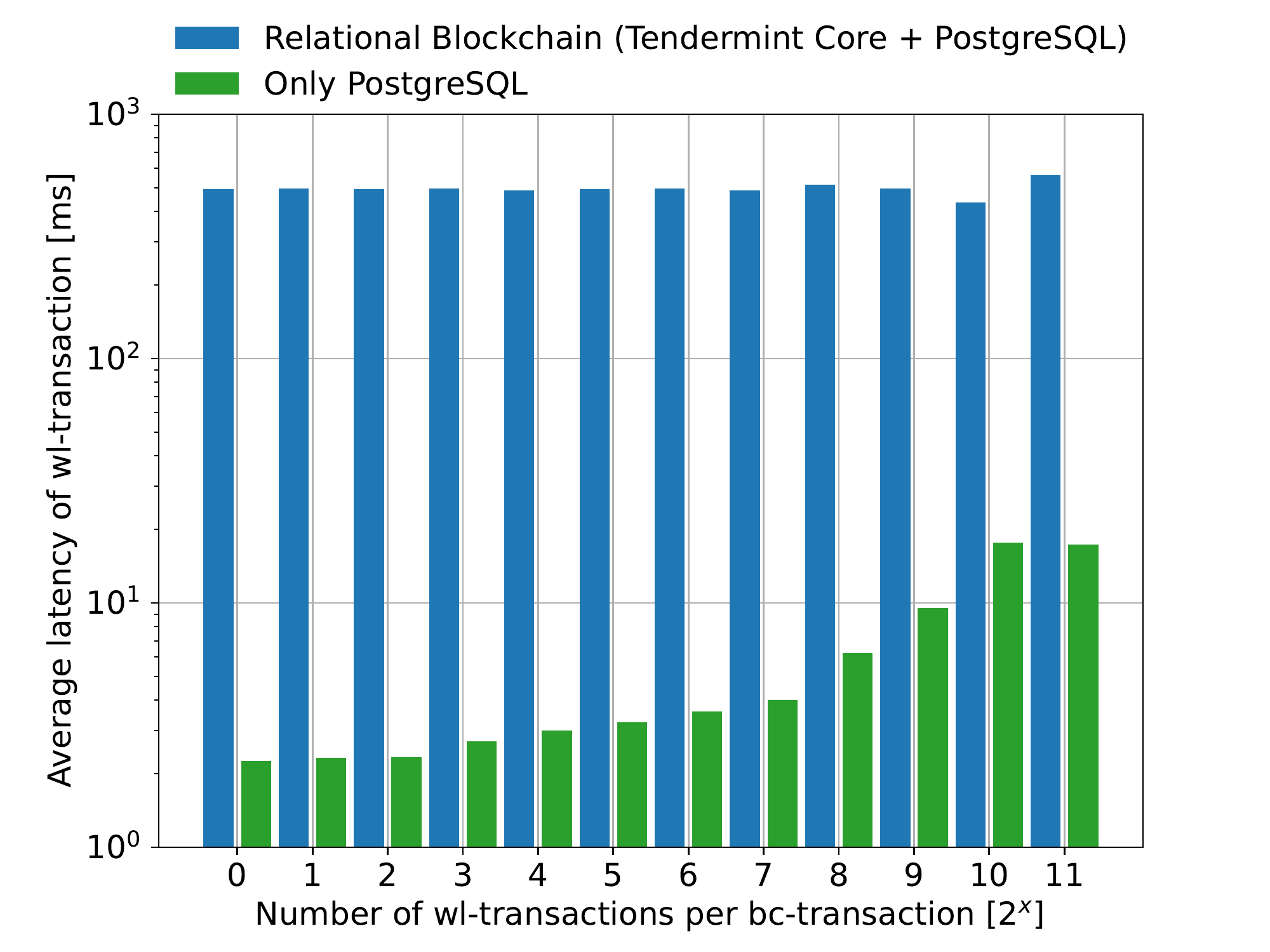}
    \caption{Latency for a mixture of writing transactions.}
    \label{fig:overhead_smb_sync_latency}
  \end{subfigure}
  \begin{subfigure}[b]{.49\textwidth}
	\includegraphics[width=\linewidth, trim={0.3cm 0 2cm 0}, clip]{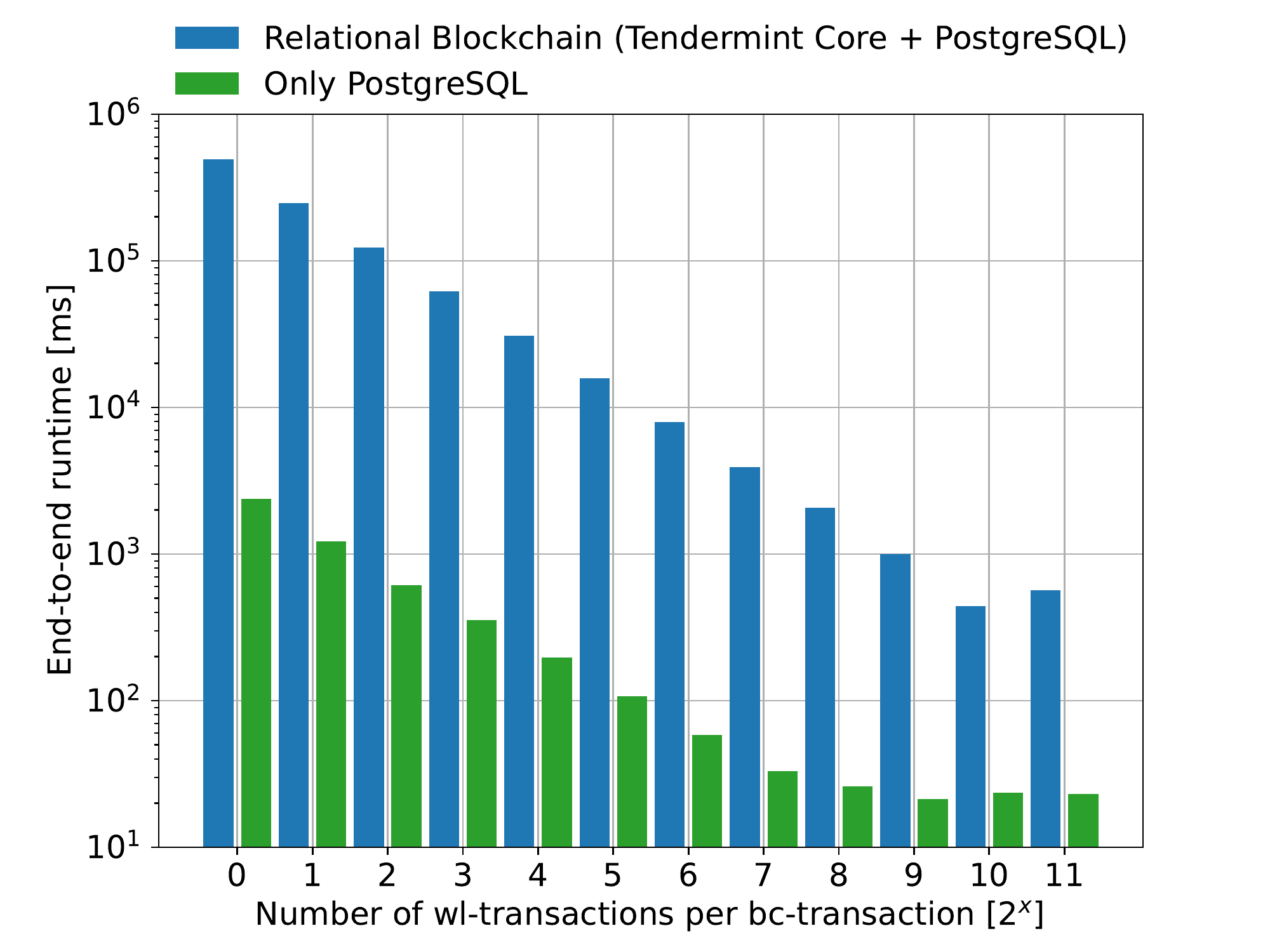}
    \caption{Throughput for a mixture of writing transactions.}
    \label{fig:overhead_smb_sync_throughput}
  \end{subfigure}
  \caption{Synchronous communication (Smallbank)}
  \label{fig:overhead_smb_sync}
\end{figure}

\begin{figure}[h!]
  \centering
  \begin{subfigure}[b]{.49\textwidth}
	\includegraphics[width=\linewidth, trim={0.3cm 0 2cm 0}, clip]{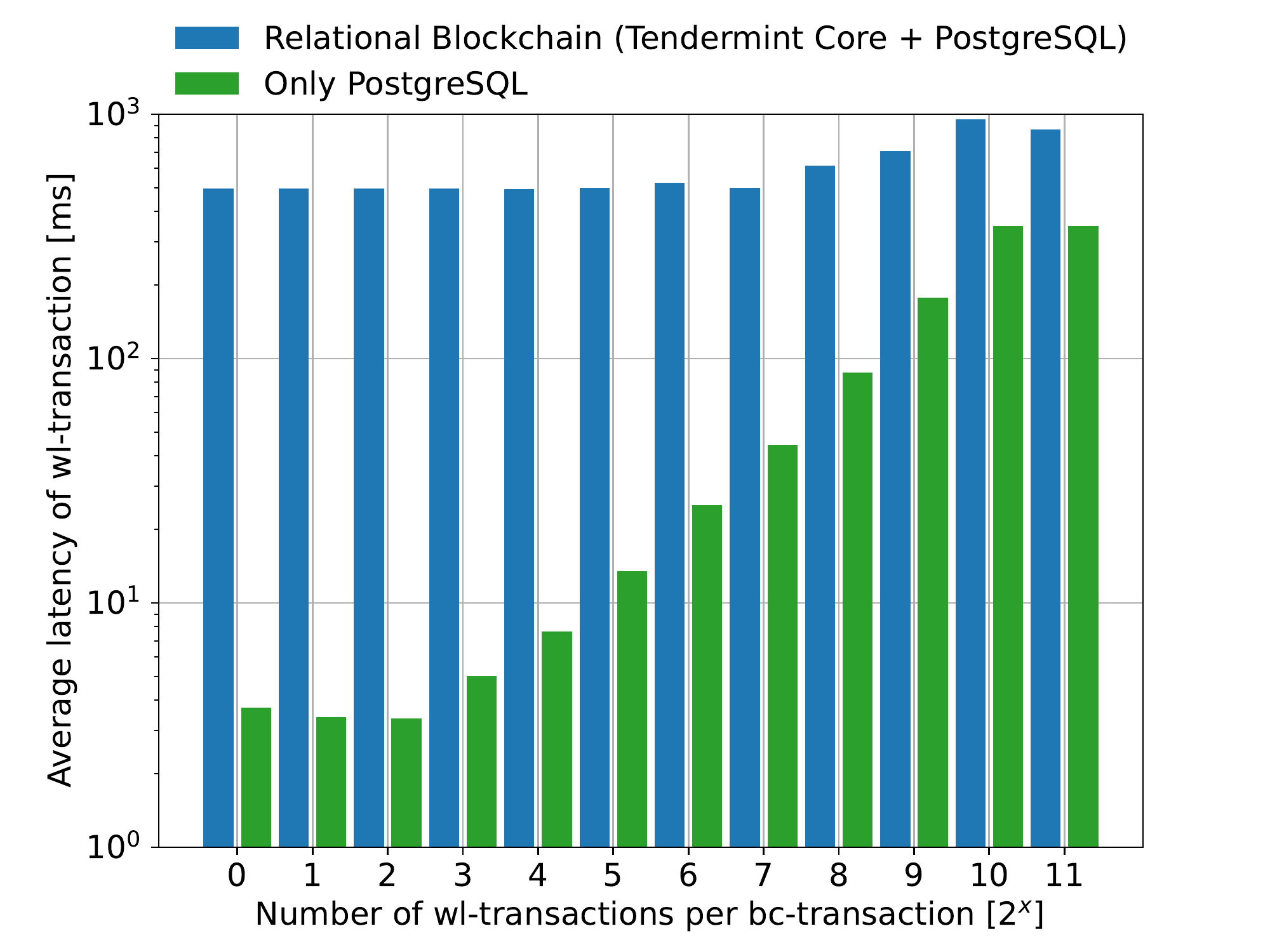}
    \caption{Latency for a mixture of writing transactions.}
    \label{fig:overhead_tpcc_sync_latency}
  \end{subfigure}
  \begin{subfigure}[b]{.49\textwidth}
	\includegraphics[width=\linewidth, trim={0.3cm 0 2cm 0}, clip]{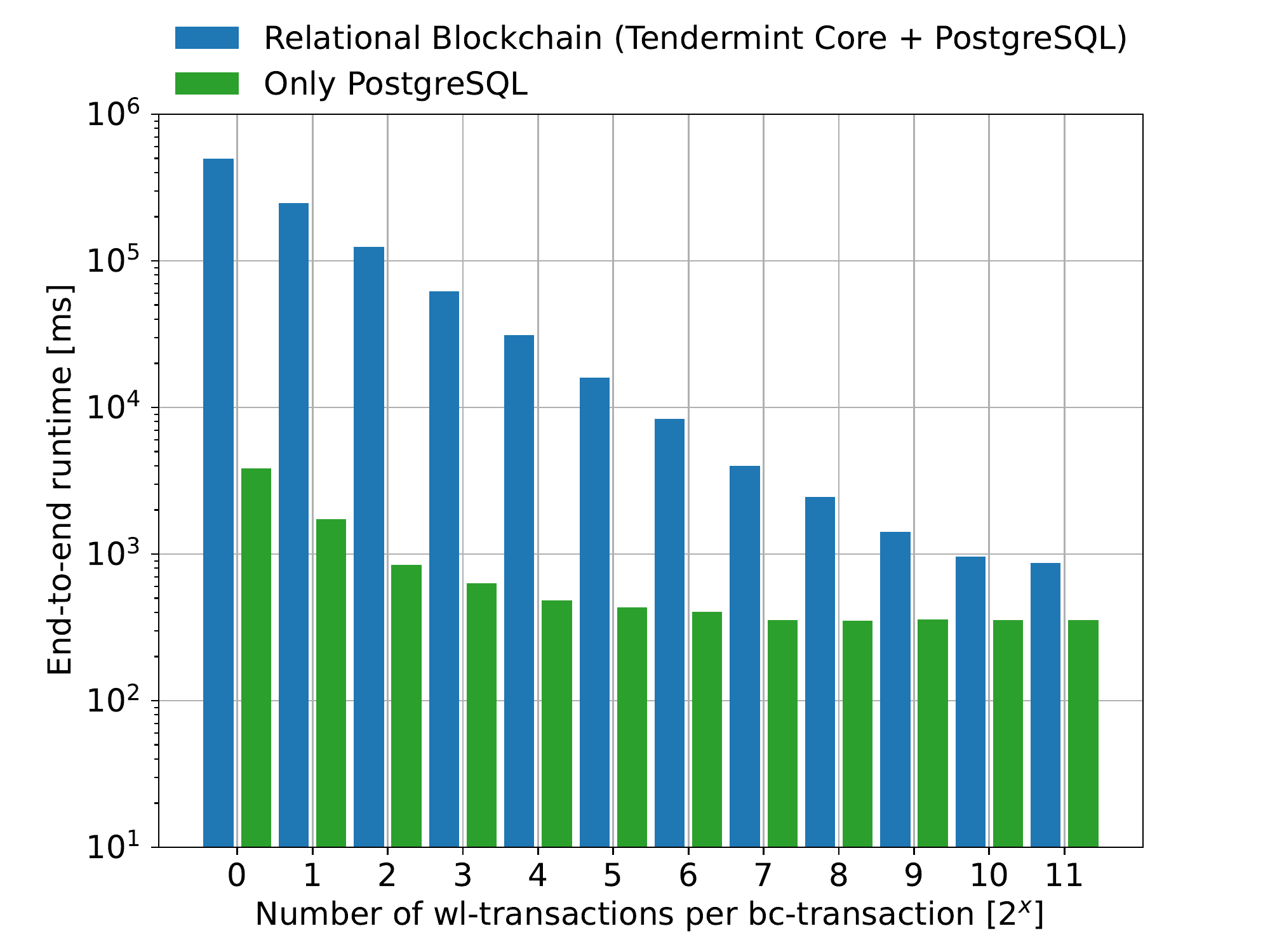}
    \caption{Throughput for a mixture of writing transactions.}
    \label{fig:overhead_tpcc_sync_throughput}
  \end{subfigure}
  \caption{Synchronous communication (TPC-C)}
  \label{fig:overhead_tpcc_sync}
\end{figure}

Figure~\ref{fig:overhead_smb_sync} and Figure~\ref{fig:overhead_tpcc_sync} show the results for Smallbank and TPC-C, respectively, where we fire a uniform mixture of $1{,}000$~writing wl-transactions in total. On the $x$-axis, we vary the number of wl-transactions per fired bc-transaction from $1$ to $2{,}048$ in logarithmic steps. As discussed, $1$ resembles the synchronous case, whereas $2$ to $2{,}048$ resemble different pseudo-synchronous configurations. 
On the $y$-axis, we show in the Figures~\ref{fig:overhead_smb_sync_latency} and \ref{fig:overhead_tpcc_sync_latency} the average latency of a wl-transaction over the whole transaction sequence. In Figures~\ref{fig:overhead_smb_sync_throughput} and \ref{fig:overhead_tpcc_sync_throughput}, we show the end-to-end runtime on the $y$-axis. To improve readability, we use a logarithmic scale on the $y$-axis.

In the results, we can observe a significant overhead of the blockchain framework over the relational DBMS under both workloads and all synchronicity configurations. However, we can also see that the overhead depends on (a)~the type of workload and (b)~the number of wl-transactions packed into a single bc-transaction, i.e., the amount of required synchronicity.

Regarding~(a), we can see that under Smallbank (Figure~\ref{fig:overhead_smb_sync}), the overhead of the blockchain framework over the raw relational DBMS is much more significant than under TPC-C (Figure~\ref{fig:overhead_tpcc_sync}). While for Smallbank, the smallest observed overhead is a still a slowdown of $32$x and $25$x in latency and end-to-end runtime, respectively, for TPC-C, the latency and runtime overhead decreases to only $2.5$x and $2.4$x in the best case. 
The reason for this lies in the complexity and individual runtime of the wl-transactions. As complex TPC-C transactions require more processing time in the relational backend than the short-running Smallbank transactions, the overhead of the framework makes a smaller fraction of the total runtime.

Regarding~(b), we observe that packing mutliple wl-transactions into a single bc-transaction heavily impacts the performance, both for the relational blockchain and the raw relational DBMS. While for the fully synchronous case, we measure a devastating overhead of $218$x (latency) and $208$x (end-to-end runtime) for Smallbank and $133$x (latency) and $129$x (end-to-end runtime) for TPC-C, the situation gradually improves when relaxing the required synchronicity. 
Particularly for TPC-C, a reduction in synchronicity has a positive impact on the amount of overhead introduced by the framework, which decreases to the aforementioned acceptable $2.5$x (latency) and $2.4$x (end-to-end runtime). This is due to the fact that less blocks are formed for the transaction sequence, requiring less consensus rounds, decreasing the central bottleneck of the blockchain framework.

\subsubsection{Asynchronous Communication}
\label{ssec:asynchronous}

Let us now look at the asynchronous case, which resembles the typical communication style with a blockchain system: The client submits a bc-transaction and the submission returns immediately. Then, after some time, the client checks whether the transaction has been bc-committed or not (yet). 

Here, ensuring a fair experimental setup between the relational blockchain and the standalone DBMS is a bit more complicated. The reason lies in the way Tendermint Core handles asynchronous transaction processing internally:
As we do not have to wait for a response, we push the whole batch of wl-transactions into the network in one go. Tendermint Core then forms blocks out of pending wl-transactions and commits them one after the other. As previously described, for each block, an individual db-transaction is opened and eventually committed, each containing a sequence of applied wl-transactions. However, as Tendermint now decides by itself how many wl-transactions it packs into a single block, it is more difficult to set up a comparable run for the standalone DBMS. To solve the problem, we record the transactions that were packed in each committed block during the run of the relational blockchain. Then, to set up the run with standalone PostgreSQL, we pack the exact same transaction sequences in individual db-transactions and fire them one after the other. 

\begin{figure}[h!]
  \centering
  \begin{subfigure}[b]{.49\textwidth}
	\includegraphics[width=\linewidth, trim={0.1cm 0 0cm 0}, clip]{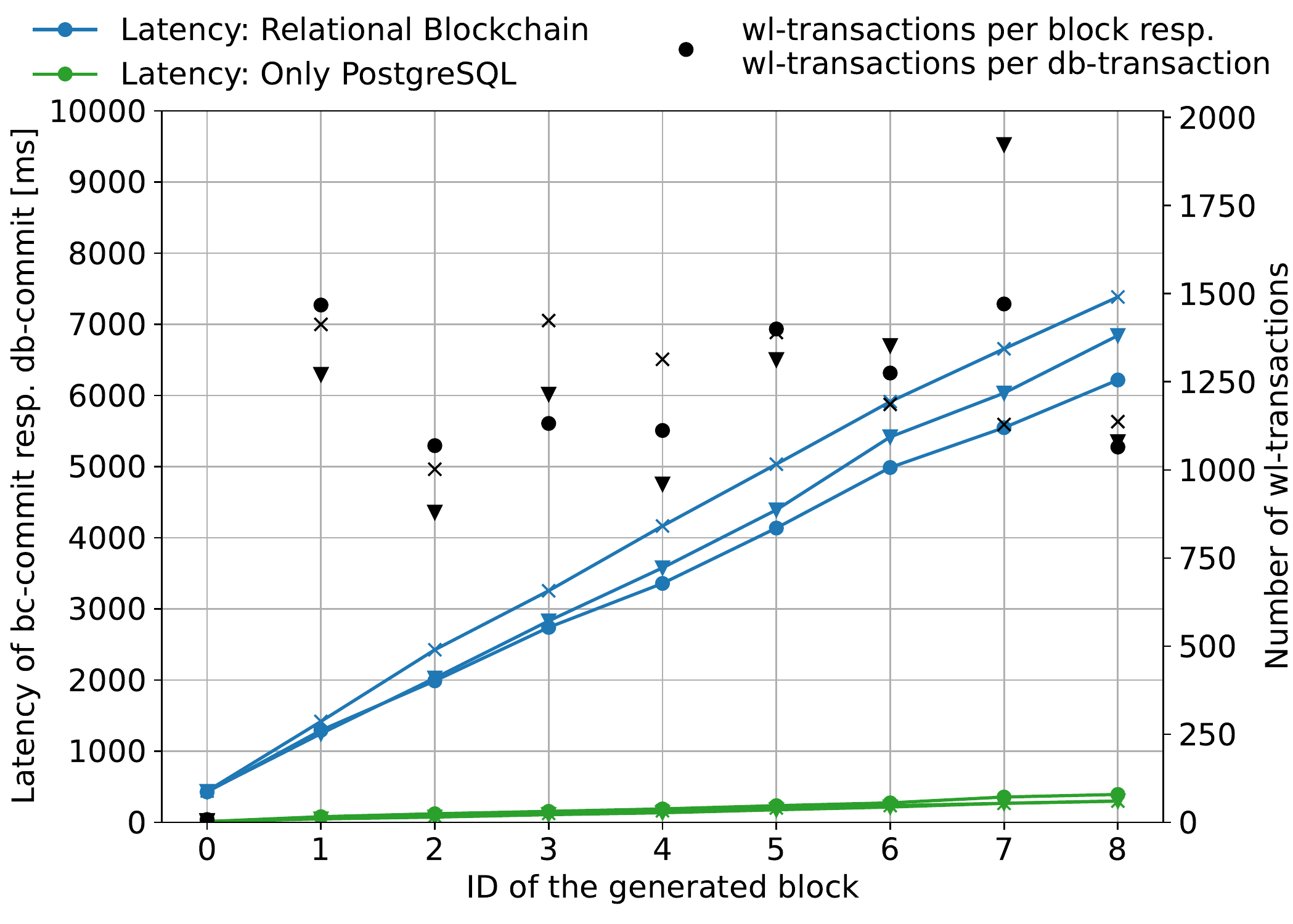}
    \caption{Latency for a mixture of writing transactions.}
    \label{fig:overhead_smb_async_latency}
  \end{subfigure}
  \hspace*{0.1cm}
  \begin{subfigure}[b]{.49\textwidth}
	\includegraphics[width=.87\linewidth, trim={0.3cm 0 1.5cm 0}, clip]{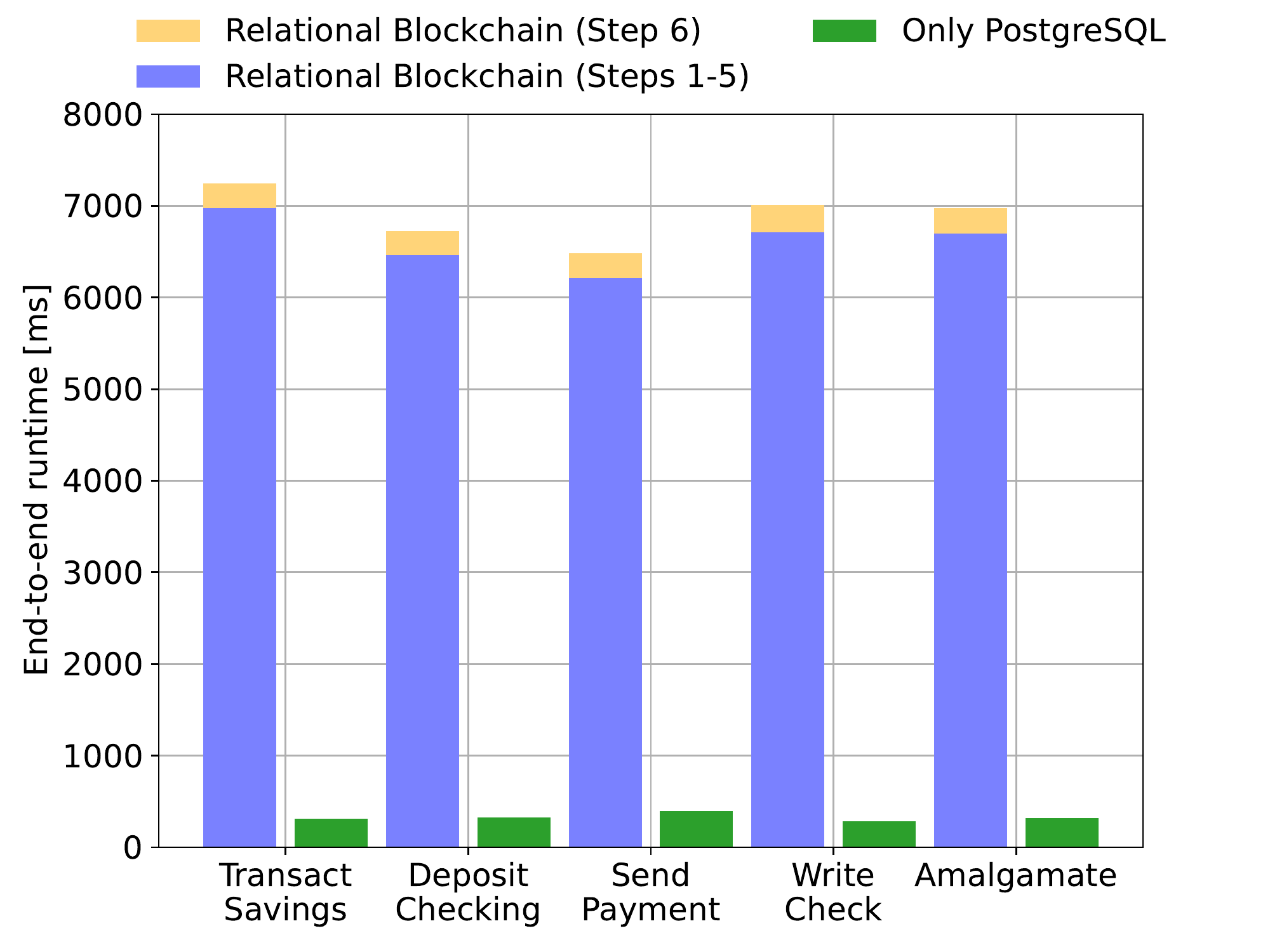}
    \caption{End-to-end runtime.}
    \label{fig:overhead_smb_async_throughput}
  \end{subfigure}
  \caption{Asynchronous communication (Smallbank)}
  \label{fig:overhead_smb_async}
\end{figure}

\begin{figure}[h!]
  \centering
  \begin{subfigure}[b]{.49\textwidth}
	\includegraphics[width=\linewidth, trim={0.1cm 0 0cm 0}, clip]{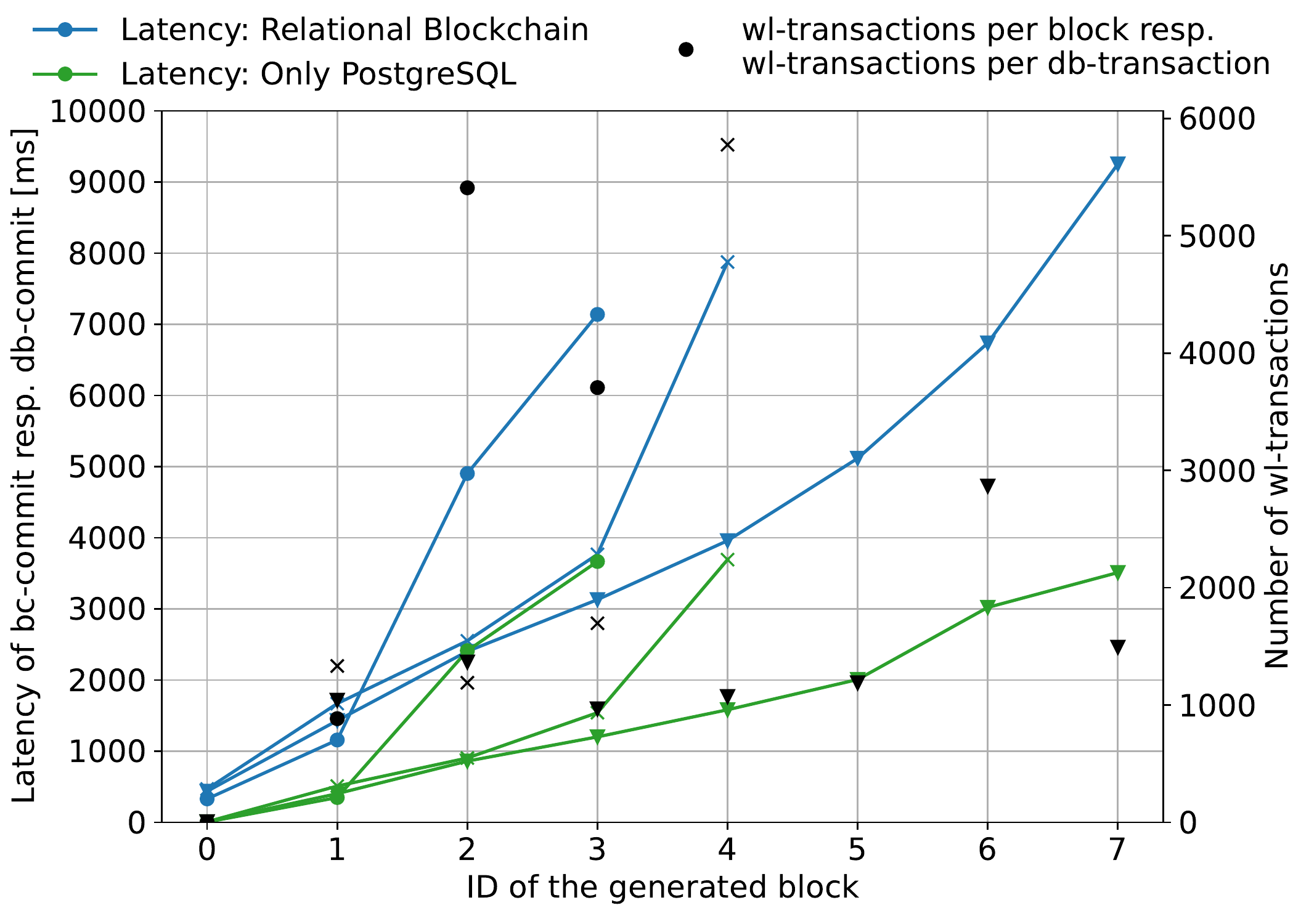}
    \caption{Latency for a mixture of writing transactions.}
    \label{fig:overhead_tpcc_async_latency}
  \end{subfigure}
    \hspace*{0.1cm}
  \begin{subfigure}[b]{.49\textwidth}
	\includegraphics[width=.94\linewidth, trim={0.1cm 0 0.3cm 0}, clip]{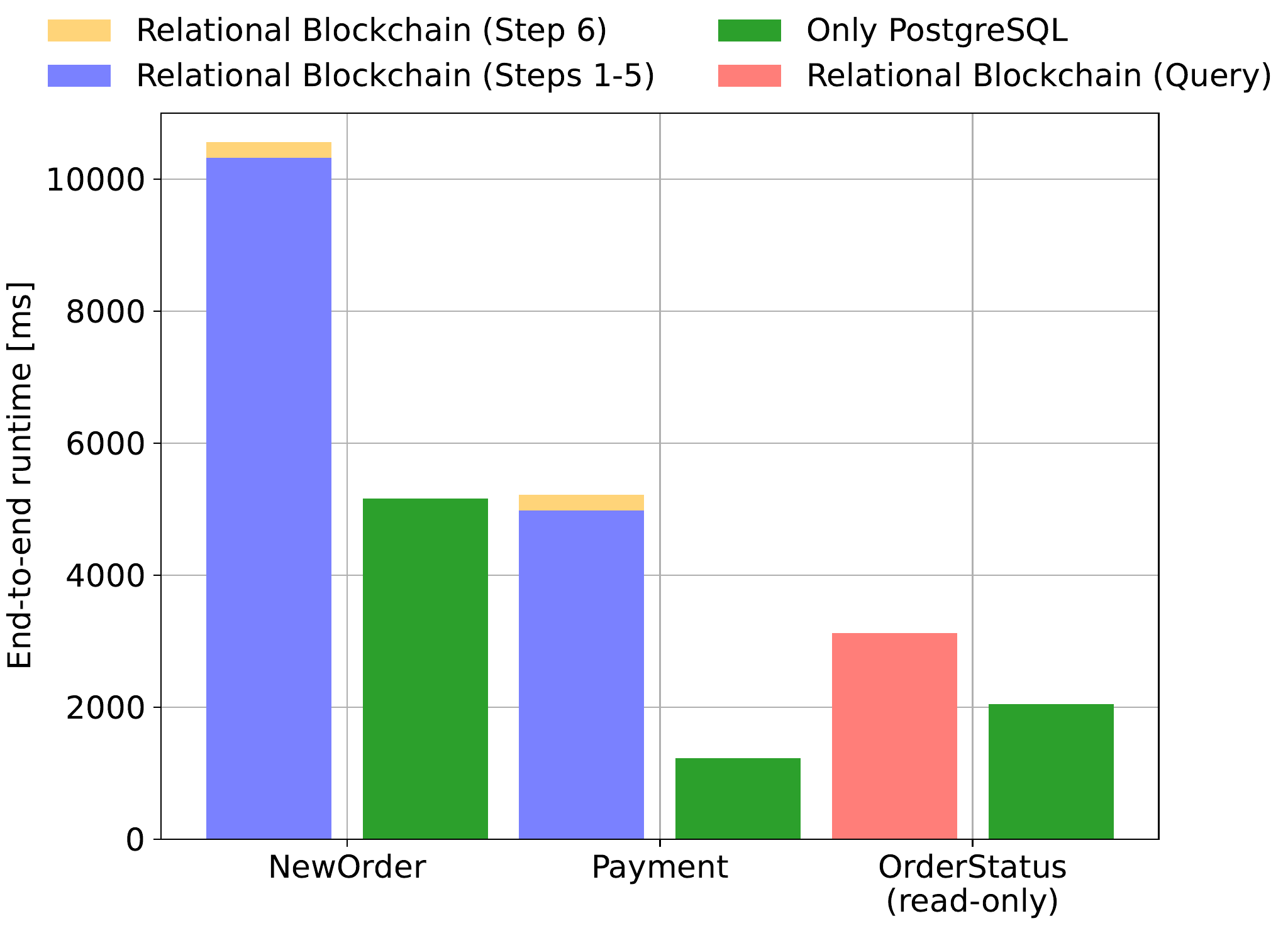}
    \caption{End-to-end runtime.}
    \label{fig:overhead_tpcc_async_throughput}
  \end{subfigure}
  \caption{Asynchronous communication (TPC-C)}
  \label{fig:overhead_tpcc_async}
\end{figure}

Figure~\ref{fig:overhead_smb_async} and Figure~\ref{fig:overhead_tpcc_async} show the experimental results. As asynchronous communication is less expensive than synchronous communication in total, we fire a larger sequence of $10{,}000$~wl-transactions this time.

In the Figures~\ref{fig:overhead_smb_async_latency} and \ref{fig:overhead_tpcc_async_latency}, we show the measured \textit{latency} of each block respectively fired db-transaction. Here, we measure latency as the time between the start of the experiment (firing the first wl-transaction) until the notification about the bc-commit of the respective block (step~4 in Figure~\ref{fig:async_test_suite}). We fire a uniformly selected mix of only writing transactions of the respective benchmark and plot the ID of each block that has been generated on the $x$-axis in relation to the latency of the corresponding bc-commit on the $y$-axis. Additionally, we plot the number of wl-transactions that were packed by the framework in each individual block (blacks dots) with respect to a second $y$-axis. As for individual runs, the framework might produce a different number of blocks, we plot each of the three performed runs individually.  

Additionally, in the Figures~\ref{fig:overhead_smb_async_throughput} and \ref{fig:overhead_tpcc_async_throughput}, we perform a set of experiments where we measure the \textit{end-to-end runtime}. In this case, we fire 10{,}000 transactions of each type individually to analyze an effect of the transaction type. For the relational blockchain, we split the end-to-end runtime for the sequence of modifying transactions into the actual transaction processing time (steps~1~to~5 of Figure~\ref{fig:async_test_suite}) and the time to check whether the transaction has been processed successfully (step~6 of Figure~\ref{fig:async_test_suite}). For the read-only transaction \texttt{OrderStatus} of TPC-C, we show the runtime when using the query-interface of the framework. 

Let us first look at the results for Smallbank in Figure~\ref{fig:overhead_smb_async}. We can see that the difference in latency and end-to-end runtime between the relational blockchain and stand-alone PostgreSQL is significant. Processing the transactions in the framework increases the latency of the last generated block respectively db-commit by up to $24$x and the end-to-end runtime by an average of $21$x over all transactions.  
We see that the result inspection (step 6) is not responsible for the overhead, the actual transaction processing in the framework takes the majority of time. We can also see that the framework packs around $1{},000$ to $2{,}000$~wl-transactions in one block, leading to the generation of $8$~blocks in total. This clearly improves the performance over the synchronous case, however, still generates significant overhead. Between the individual transaction, we observe little difference. All transactions are extremely short-running in the backend and modify at most two accounts each.

Let us now inspect the TPC-C results in Figure~\ref{fig:overhead_tpcc_async}, which look quite different to the results of Smallbank. First of all, we can see that the overhead of the framework over stand-alone PostgreSQL is significantly smaller for this benchmark. This time, the framework increases the latency at most by $2.6$x. The end-to-end runtime of the sequence of \texttt{NewOrder} and \texttt{Payment} transactions increases only by $2.0$x and $4.3$x, respectively. The reason lies in the much higher complexity of the performed transaction: If the backend requires more time to process a transaction, the overhead of processing it in the framework becomes less significant in the end-to-end runtime. 
For the read-only transaction, the overhead of the framework is even smaller with $1.52$x, as we can bypass block forming and consensus entirely. This shows that for queries, the framework should be bypassed entirely. 

Overall, we can also see that the overhead under asynchronous communication is significantly smaller than in the synchronous case. 

\subsection{Impact of the Relational Backend}
\label{ssec:dbms}

\begin{figure}[h!]
  \centering
  \begin{subfigure}[b]{.49\textwidth}
	\includegraphics[width=\linewidth, trim={0 0 2cm 0}, clip]{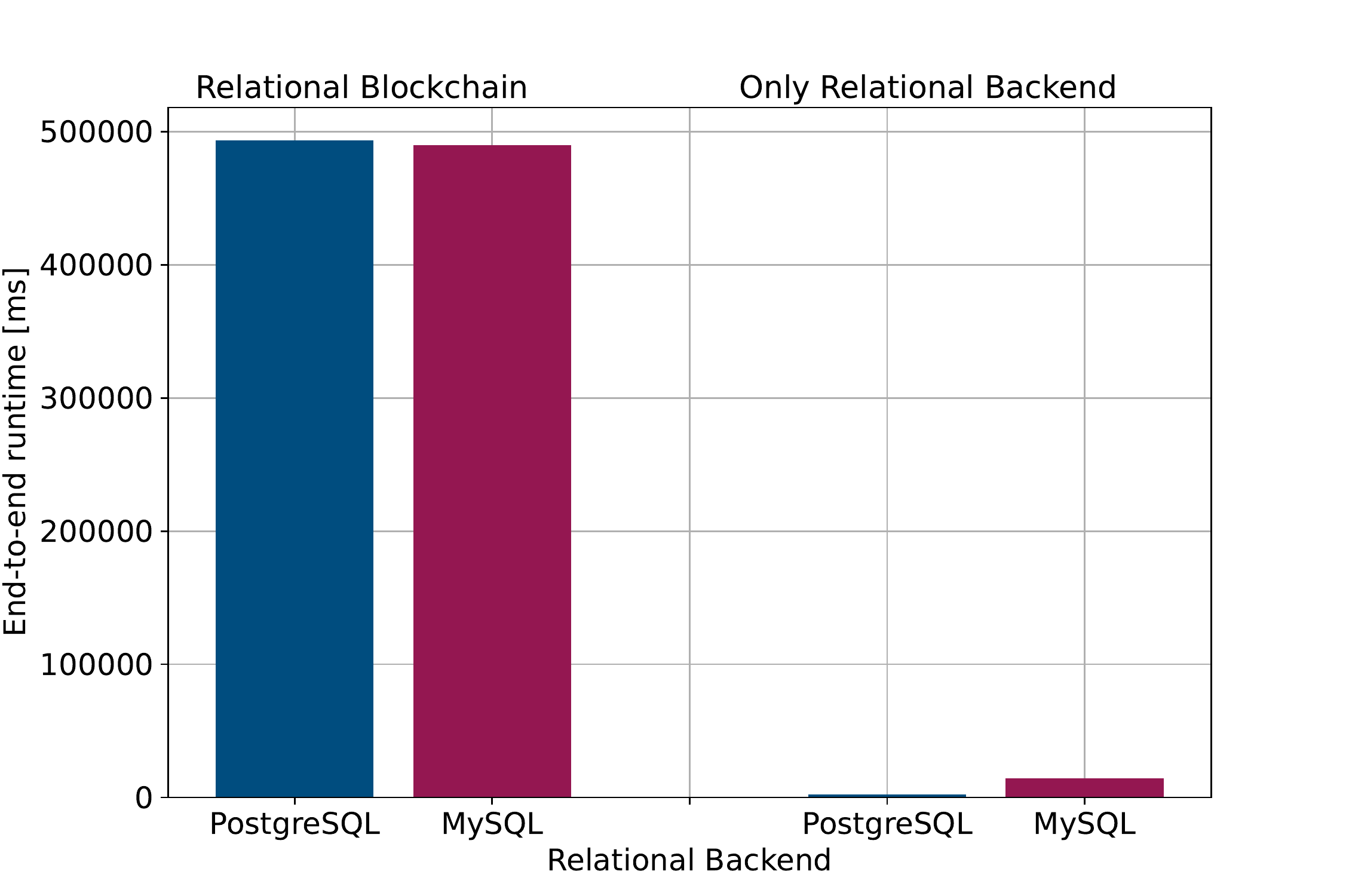}
    \caption{Synchronous communication.}
    \label{fig:dbms_backend_sync}
  \end{subfigure}
  \begin{subfigure}[b]{.49\textwidth}
	\includegraphics[width=.9\linewidth, trim={0 0 1.5cm 0}, clip]{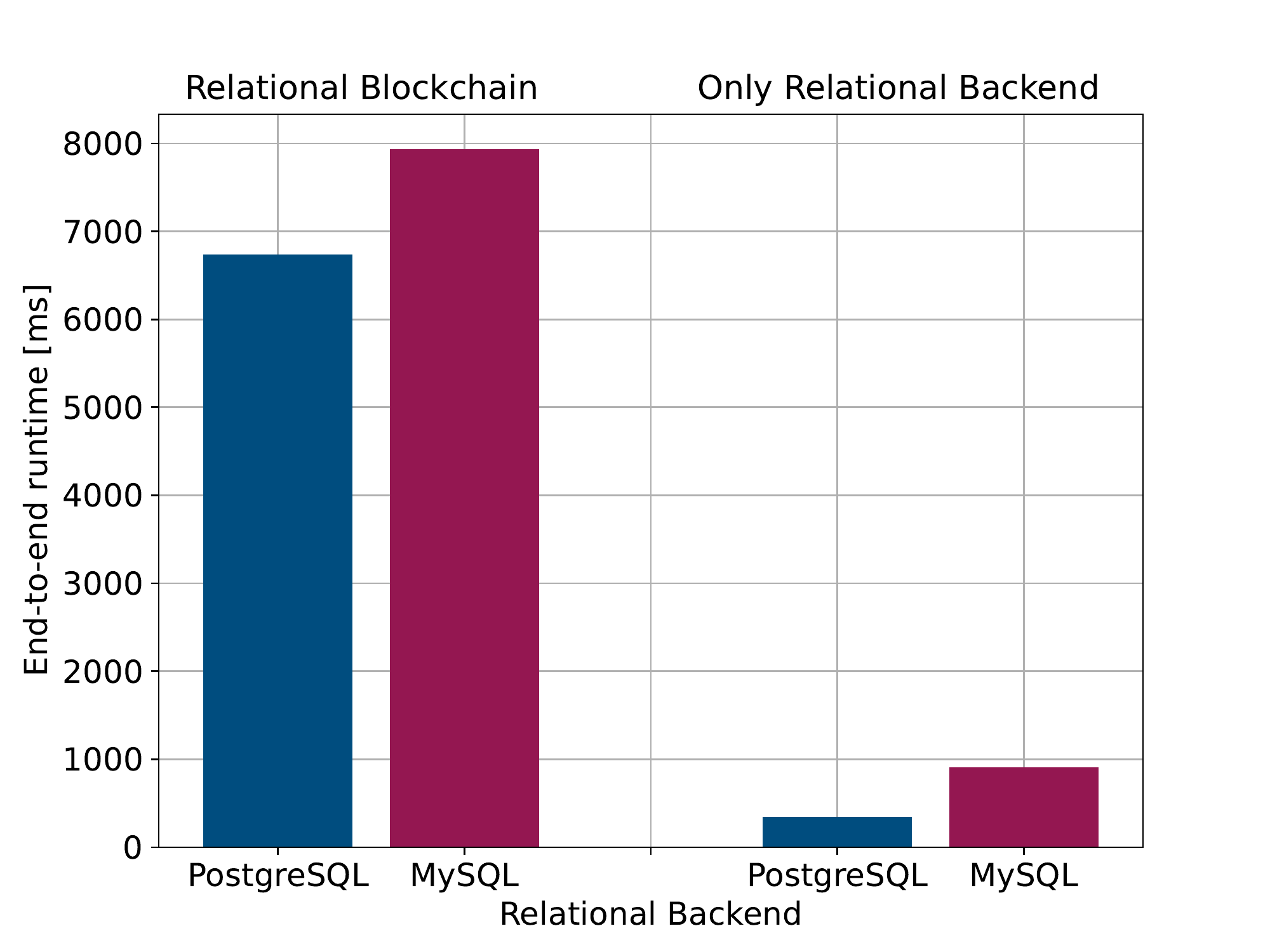}
    \caption{Asynchronous communication.}
    \label{fig:dbms_backend_async}
  \end{subfigure}
  \caption{Impact of the relational DBMS in the backend.}
  \label{fig:dbms_backend}
\end{figure}

So far, we used PostgreSQL as the relational DBMS in the backend for all experiments. Let us now investigate whether the choice of the relational system actually matters or whether its performance is completely overshadowed, if it is embedded in the blockchain framework. 
In Figure~\ref{fig:dbms_backend}, we show the end-to-end runtime of our relational blockchain under a uniform mixture of $1{,}000$~synchronous respectively $10{,}000$~asynchronous wl-transactions of the Smallbank workload, where we use either PostgreSQL or MySQL as the backend in all four virtual nodes. 

Let us first look at the raw backend performance shown on the right side of the plots. We can see that PostgreSQL is able to process the sequence of transactions significantly faster than MySQL. In the synchronous case, PostgreSQL is $6.6$x faster than MySQL. In the asynchronous case, where multiple wl-transactions are packed in a single db-transaction, the speedup is still $2.6$x. 
While the backends perform drastically different on the workload, this difference becomes less significant when embedding the backend within the relational blockchain. In the synchronous case, the backend makes no difference at all, as the runtime is dominated by block forming and consensus. Only in the asynchronous case, we see a significant difference. Therein, using PostgreSQL improves the end-to-end runtime by $1.2$x over MySQL.

\subsection{Impact of Scaling across Virtual Nodes and Physical Nodes}
\label{ssec:scaling}

\begin{figure}[h!]
  \centering
  \begin{subfigure}[b]{.49\textwidth}
	\includegraphics[width=\linewidth, trim={0 0 0 1.1cm}, clip]{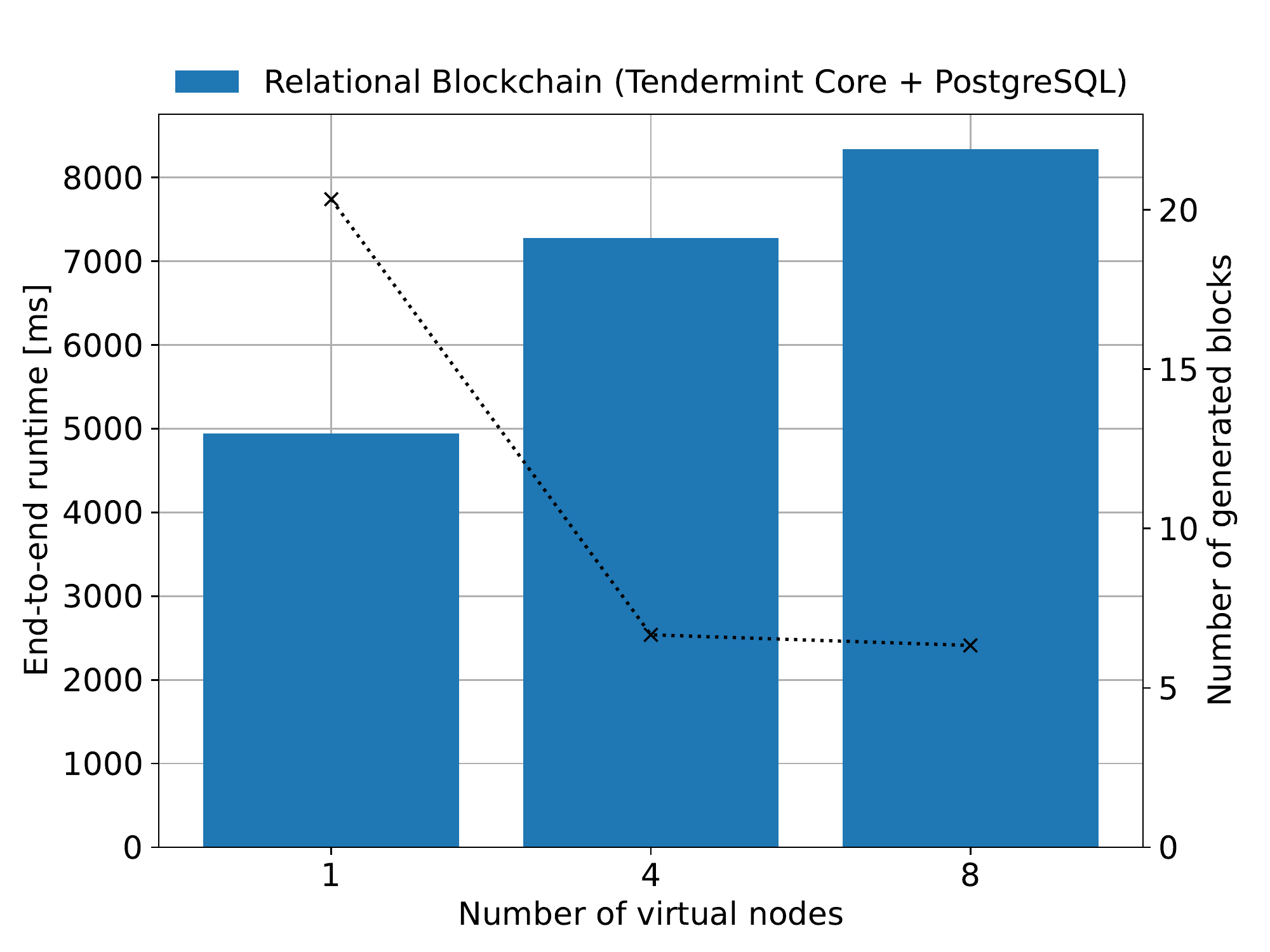}
    \caption{Virtual scaling.}
    \label{fig:virtual_scaling_smb}
  \end{subfigure}
  \begin{subfigure}[b]{.49\textwidth}
	\includegraphics[width=\linewidth, trim={0 0 0 1.1cm}, clip]{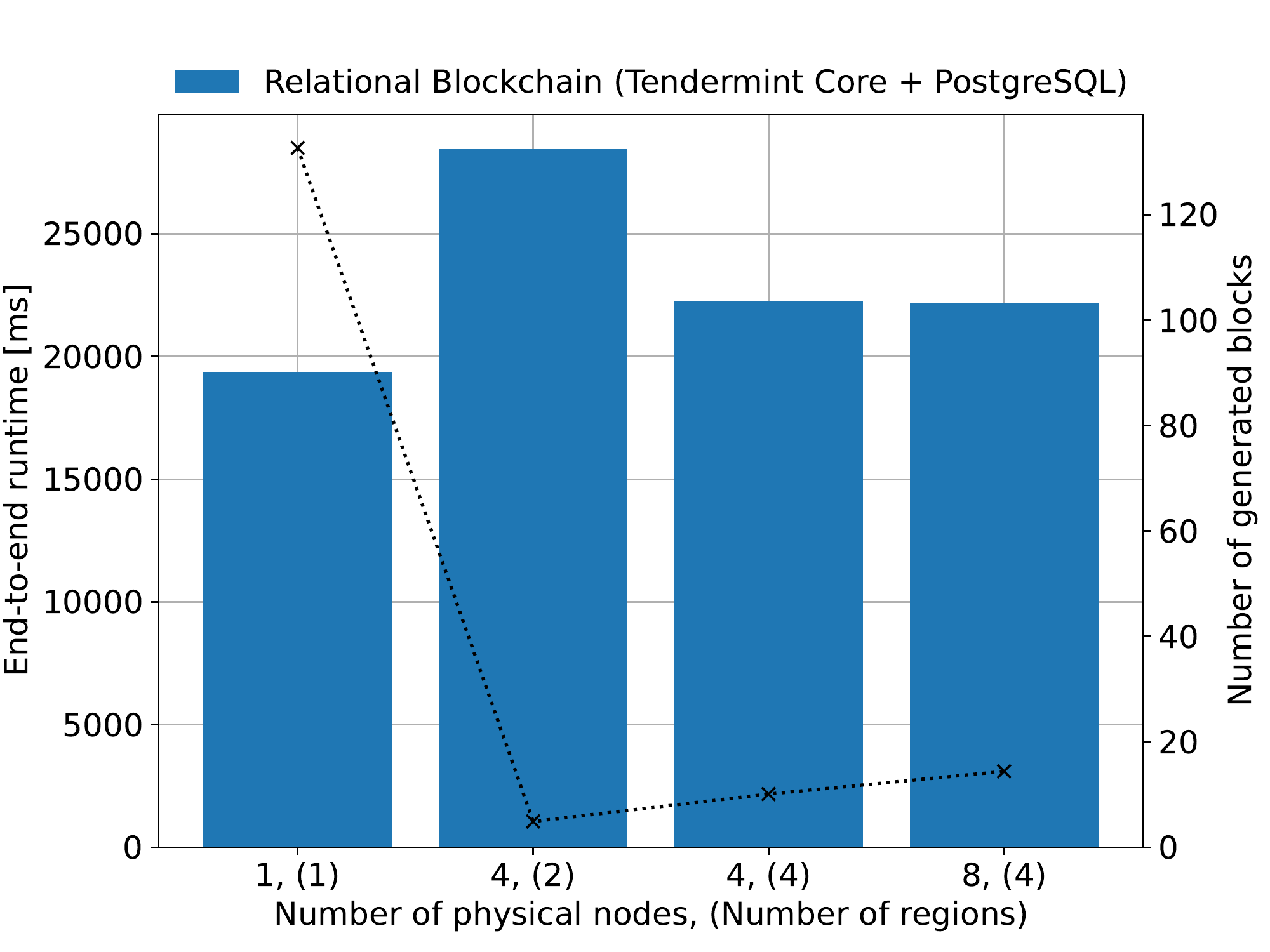}
    \caption{Physical scaling.}
    \label{fig:physical_scaling_smb}
  \end{subfigure}
  \caption{Scaling the number of virtual and physical nodes.}
  \label{fig:scaling_smb}
\end{figure}

Until now, we ran all experiments using four virtual nodes running on one physical node. In the following, we will vary both the number of virtual nodes (Figure~\ref{fig:virtual_scaling_smb}) as well as the number of physical nodes (Figure~\ref{fig:physical_scaling_smb}) to represent the network. 

We start by varying the number of virtual nodes in Figure~\ref{fig:virtual_scaling_smb}. This experiment still factors out network latency. We set up a network of only one virtual node, four~virtual nodes, and eight~virtual nodes and report the end-to-end runtime for $1{,}000$~modifying Smallbank transactions using asynchronous communication. Additionally, we show the number of generated blocks for the total run. Note that a network consisting of only one node can skip the consensus phase, as no other participants exist to coordinate with. We see this setup as the baseline for the throughput that can be achieved in the system. 
When looking at the results in Figure~\ref{fig:virtual_scaling_smb}, we can see that the end-to-end runtime is unsurprisingly the shortest when running only one node. When using four~nodes, the runtime increases by a factor of $1.47$x over the single node configuration, when using eight~nodes, it increases by a factor of $1.69$x. This shows that an increase in the number of nodes clearly increases the overhead, however, only sublinearly. When inspecting the number of generated blocks, we can see that for one node, $20$~(smaller)~blocks are generated on average, whereas for four and for eight~nodes, only $6$~(larger)~blocks are generated for the whole sequence of $1{,}000$~transactions. This shows that the consensus that is performed for a block throttles the forming of the next block. 

Let us now look at the results when scaling the number of physical nodes in Figure~\ref{fig:physical_scaling_smb}.
Here, we test one physical node in one region (Frankfurt), four physical nodes in two different regions (Frankfurt and Paris), four physical nodes in all four different regions, and eight physical nodes in all four different regions. Note that for this experiment, we repeat each run $10$~times (instead of $3$~times as before) to factor out variance caused by the cloud provider as much as possible. 
First of all, we can observe that the end-to-end runtime is overall higher than when scaling the number of virtual nodes within one physical node. This is caused by the internet latency, but also by the slower physical nodes. 
Again, using only one node is unsurprisingly fastest, however, using more physical nodes does not decrease the performance as heavily as for virtual scaling. Using four physical nodes within two regions shows worse performance than four physical nodes within four regions. We deduct from this that the internet traffic between the nodes is not the bottleneck here, but that the two physical nodes that are in the same region potentially share the same hardware resources. Going to eight physical nodes decreases the performance only marginally in comparison to four nodes. 

\section{Takeways and Conclusion}
\label{sec:conclusion}

In this work, we have presented a practical and feasible way of integrating a full-fledged relational DBMSs into a blockchain framework to support the execution of SQL transactions in byzantine environments. We analyzed the performance implications of such a systems combination and identified situations where the overhead is acceptable. Also, we have seen setups where the overhead is dramatic and completely overshadows the backend performance. In the following, we conclude with the most important findings and takeaway messages, which hopefully support users to achieve the best performance of such a setup:

\begin{itemize}
\item Independent of the faced workload and the installed backend, synchronous communication throttles the performance of the relational blockchain severely. We recommend synchronous communication only if it is absolutely required by the application. Pseudo-synchronous communication reduces the performance problems -- the more transactions are answered in a batch, the lower the latency and the higher the throughput. The best performance is observed under asynchronous communication, where the framework decides on its own how many transactions to pack into a block.

\item Under pseudo-synchronous and asynchronous communication, the type of workload has an impact on both performance and overhead. While for short-running transactions, the framework is the bottleneck, for more complex transactions, the relational backend becomes the dominant factor with an increase in block size. 

\item Read-only transactions cause significantly less overhead than modifying transactions, as they can bypass the costly transaction processing flow that modifying transactions must go through.

\item The type of backend is relevant for asynchronous communication only, not for synchronous communication. Even for simple and short-running transactions, such as in Smallbank, the backend has an impact on the runtime. For more complex transactions, the impact is likely to be more significant as well.  

\item The presented relational blockchain scales gracefully both locally with the number of virtual nodes and across data-centers with the number of physical nodes.
\end{itemize}


\bibliography{relational_tendermint} 

\end{document}